\documentclass[fleqn,usenatbib]{mnras}
\usepackage[T1]{fontenc}
\usepackage{ae,aecompl}
\usepackage{graphicx}	
\usepackage{amsmath}	
\usepackage{amssymb}	
\usepackage{tikz}
\usepackage{times}
\usepackage[normalem]{ulem}
\usepackage{multirow}
\usepackage{tabularx}
\usepackage{bm}            
\usepackage{url}
\usepackage{color,colortbl}

\usepackage{hyperref}

\pdfoutput=1

\definecolor{Gray}{gray}{0.9}

\newcommand{\pc}{\ensuremath{~\mathrm{pc}}}
\newcommand{\kpc}{\ensuremath{~\mathrm{kpc}}}
\newcommand{\Mpc}{\ensuremath{~\mathrm{Mpc}}}
\newcommand{\kms}{\ensuremath{~\mathrm{km~s^{-1}}}}

\newcommand{\Msun}{\ensuremath{~\mathrm{M}_\odot}}

\newcommand{\Mbar}{\ensuremath{M_{\mathrm{bar}}}}
\newcommand{\Mtot}{\ensuremath{M_{\mathrm{tot}}}}
\newcommand{\MtotDMO}{\ensuremath{M_{\mathrm{tot}}^{\mathrm{DMO}}}}
\newcommand{\Mdm}{\ensuremath{M_{\mathrm{DM}}}}
\newcommand{\MdmDMO}{\ensuremath{M_{\mathrm{DM}}^{\mathrm{DMO}}}}
\newcommand{\MdmPred}{\ensuremath{M_{\mathrm{DM}}^{\mathrm{pred}}}}
\newcommand{\fbar}{\ensuremath{f_{\mathrm{bar}}}}
\newcommand{\Vcirc}{\ensuremath{V_{\mathrm{circ}}}}

\newcommand{\HI}{H\textsc{i}}
\newcommand{\HII}{H$_2$}

\newcommand{\figDir}{fig_pdf/}

\newcommand{\eagle}{EAGLE\_recal}
\newcommand{\eagleS}{EAGLE}
\newcommand{\auriga}{Auriga}
\newcommand{\apostle}{APOSTLE}

\usepackage{color}
\definecolor{mycolor}{rgb}{.0,.3,1.}
\newcommand{\changed}[1]{{#1}}

\begin{document}



\title[The MW mass profile]{The Milky Way total mass profile as inferred from \textit{Gaia} DR2}

\author[Cautun et al.]
{\parbox{\textwidth}{
    Marius~Cautun$^{1,2}$\thanks{E-mail: cautun@strw.leidenuniv.nl}, 
    Alejandro Ben\'{i}tez-Llambay$^{2}$, 
    Alis~J.~Deason$^{2}$, 
    Carlos~S.~Frenk$^2$,
    Azadeh Fattahi$^2$,
    Facundo A. G\'{o}mez$^{3,4}$,
    Robert J. J. Grand$^{5}$,
    Kyle A. Oman$^{2}$, \linebreak
    Julio F. Navarro$^{6}$ and
    Christine M. Simpson$^{7,8}$
    }
\vspace{.2cm}\\
$^{1}$ Leiden Observatory, Leiden University, PO Box 9513, NL-2300 RA Leiden, the Netherlands \\
$^{2}$ Institute of Computational Cosmology, Department of Physics, Durham University, South Road, Durham, DH1 3LE, UK \\
$^3$Instituto de Investigaci{\'o}n Multidisciplinar en Ciencia yTecnolog{\'i}a, Universidad de La Serena, Ra{\'u}l Bitr{\'a}n 1305, La Serena, Chile\\
$^4$Departamento de F{\'i}sica y Astronom{\'i}a, Universidad de LaSerena, Av. Juan Cisternas 1200 N, La Serena, Chile\\
$^5$Max-Planck-Institut f\"{u}r Astrophysik, Karl-Schwarzschild-Str. 1, 85748 Garching, Germany\\
$^{6}$ Department of Physics \& Astronomy, University of Victoria, Victoria, BC, V8P 5C2, Canada\\
$^7$Enrico Fermi Institute, The University of Chicago, Chicago, IL 60637, USA\\
$^8$Department of Astronomy \& Astrophysics, The University of Chicago, Chicago, IL 60637, USA\\
}


\pubyear{2019}

\label{firstpage}
\pagerange{\pageref{firstpage}--\pageref{lastpage}}
\maketitle

\begin{abstract}
  We determine the Milky Way (MW) mass profile inferred from fitting
physically motivated models to the \textit{Gaia} DR2 Galactic rotation
curve and other data. Using various hydrodynamical simulations of MW-mass
haloes, we show that the presence of baryons induces a contraction of
the dark matter (DM) distribution in the inner regions, $r\lesssim 20$
kpc. We provide an analytic expression that relates the baryonic distribution to the change in the DM halo profile. For our galaxy, the contraction increases the enclosed DM halo mass by factors of
roughly $1.3$, $2$ and $4$ at radial distances of 20, 8 and 1 kpc,
respectively compared to an uncontracted halo.  Ignoring this
contraction results in systematic biases in the inferred halo mass and
concentration. We provide a best-fitting contracted NFW halo model to
the MW rotation curve that matches the data very well\footnotemark.
The best-fit
has a DM halo mass,
$M_{200}^{\rm DM}=0.97_{-0.19}^{+0.24}\times10^{12}\Msun$, and
concentration before baryon contraction of $9.4_{-2.6}^{+1.9}$, which
lie close to the median halo mass--concentration relation
predicted in $\Lambda$CDM. The inferred total mass,
$M_{200}^{\rm total}=1.08_{-0.14}^{+0.20} \times 10^{12}\Msun$, is in
good agreement with  recent measurements. The model gives a MW
stellar mass of $5.04_{-0.52}^{+0.43}\times10^{10}\Msun$ and infers that the DM density at the Solar position is
$\rho_{\odot}^{\rm DM}=8.8_{-0.5}^{+0.5}\times10^{-3}\Msun\pc^{-3}\equiv0.33_{-0.02}^{+0.02}~\rm{GeV}~\rm{cm}^{-3}$.  The rotation
curve data can also be fitted with an uncontracted NFW halo model, but
with very different DM and stellar parameters. The observations prefer
the physically motivated contracted NFW halo, but the measurement
uncertainties are too large to rule out the uncontracted NFW halo.
\end{abstract}


\footnotetext{The 
data products are publicly available in \citet{cautun_2020_3740067} at: \href{https://github.com/MariusCautun/Milky_Way_mass_profile}{https://github.com/MariusCautun/Milky\_Way\_mass\_profile}
}

\section{Introduction} 
\label{sec:introduction}

The wealth of data available for the Milky Way (MW) makes our galaxy
an unmatched laboratory for testing cosmology on the smallest scales
and for understanding galaxy formation physics in detail
\citep[e.g. see the reviews by][]{Bullock2017,Zavala2019}. The results
of many of these tests are sensitive to the dark matter (DM) content
of our galaxy and, in particular, to the total mass and the radial
density profile of our Galactic halo. For example, the total number of
subhaloes is very sensitive to the host halo mass
\citep[e.g.][]{Purcell2012,Wang2012j,Cautun2014b,Hellwing2016} while
the radial mass profile plays a key role in determining the orbits of
satellite galaxies and tidal streams
\citep[e.g.][]{Barber2014,Patel2017,Fritz2018,Cautun2019a,Garavito-Camargo2019}. The
number and orbits of satellites are a key test of properties of the
DM, such as the mass of the DM particle and its interaction
cross-section
\citep[e.g.][]{Penarrubia2010,Vogelsberger2012,Kennedy2014,Lovell2014,Cautun2017,Kahlhoefer2019},
and also constrain galaxy formation models
\citep[e.g.][]{Sawala2016b,Sawala2016a,Bose2018,Shao2018a,Fillingham2019}.

Most previous studies have focused on determining the total mass of
the Galactic DM halo using a variety of methods, such as the dynamics
of the stellar halo \citep[e.g.][]{Xue2008,Deason2012,Kafle2012},
globular clusters \citep[e.g.][]{Eadie2016,Posti2018,Watkins2019} and
satellite galaxies
\citep[e.g.][]{Watkins2010,Li2017,Patel2017b,Callingham2019}, high velocity
stars \citep[e,g,][]{Smith2007,
  Piffl2014,Fragione2017,Rossi2017,Deason2019}, the orbits of tidal
streams \citep[e.g.][]{Gibbons2014,Bowden2015}, the luminosity
function of the MW satellites \citep[e.g.][]{Busha2011a,Cautun2014c}
and the dynamics of the Local Group \citep[e.g.][]{Li2008, Diaz2014,
  Penarrubia2016}. However, recent estimates of the total mass of the
MW still range within about a factor of two \citep[see e.g. Figure 7
in][]{Callingham2019}, reflecting systematics in many of the methods
used to infer it \citep[e..g][]{Wang2015a,Wang2017,Wang2018}.

The radial density profile of the MW is even more poorly 
\changed{measured}
due to a lack of data outside ${\sim}20\kpc$ and uncertainties in
modelling the effect of baryons on the DM halo. Most studies assume
that the DM halo is well described by an NFW profile
\citep{Navarro1996,Navarro1997} and constrain the profile by 
two parameters, such as total mass and concentration
\citep[e.g.][]{McMillan2011,Bovy2012a,Eilers2019}. Such studies argue
that the Galactic halo has a very high concentration, typically
${\sim}14$ or higher \citep[e.g.][]{Deason2012,
  Kafle2014,McMillan2017,Monari2018,Lin2019}, that is in tension with
theoretical expectations based on cosmological simulations, which
predict a mean concentration of ${\sim}9$ and a $68$ percentile range
of ${\sim}[7,12]$ \citep{Ludlow2014,Hellwing2016,Klypin2016}.

The higher than expected concentration of the MW halo could be a
manifestation of the contraction of the DM halo induced by the
presence of a galaxy at its centre
\citep[e.g.][]{Schaller2015a,Dutton2016,Lovell2018a}. For MW and
higher mass haloes, the effect of baryons on the DM halo is well
described by the adiabatic contraction model \citep{Callingham2019b},
in which baryons slowly accumulate at the halo centre and the DM
distribution distorts in such a way that its action
integrals remain approximately constant \citep{Barnes1984,Blumenthal1986,Barnes1987}. This
process can be implemented analytically if the distribution of DM
actions in the absence of baryons is known \citep{Sellwood2005};
however, since this is not well known and there is halo-to-halo
variation, in practice most studies have used approximations of this
process \citep[e.g. see][]{Blumenthal1986,Gnedin2010,Abadi2010}. Such approaches have
only occasionally been used when analysing MW data
\citep[e.g.][]{Piffl2015,Cole2017}, and most studies ignore the change
in the DM profile induced by the condensation of baryons at the centre
of haloes, despite, as we shall see, the fact that it is a large
effect, especially in the inner $10\kpc$ of our galaxy.

In this paper, we provide a best fitting mass model for the MW using
the latest Gaia rotation curve \citep{Eilers2019} combined with the
robust and extensively tested total mass determination of
\citet{Callingham2019}. We improve on previous studies by modelling
the contraction of the DM halo induced by the central galaxy. We study
the DM halo contraction and propose a simple parametric model based on
the predictions of three state-of-the-art galaxy formation
simulations: \auriga{} \citep{Grand2017}, \apostle{}
\citep{Fattahi2016,Sawala2016a} and \eagleS{} \citep{Schaye2015}, and find that all
three simulations predict the same DM halo contraction within the
limits of halo-to-halo variation. We show that the contracted DM halo
cannot be modelled as a pure NFW profile and even more flexible formulae,
such as the generalised NFW profile (gNFW, which has been used to
model the MW halo -- \citealt{McMillan2017,Karukes2019}), struggle to
describe the radial profile of the contracted halo.

We model the MW galaxy using seven components \citep[similar to the
approach used by][]{McMillan2017}: a bulge, a thin and a thick stellar disc, an
\HI{} and a molecular gas disc, a circumgalactic medium (CGM)
component, and a DM halo. Our main results are for a DM halo that has
been contracted according to the self-consistently determined MW
stellar mass. For comparison, we use a second model in which the DM
halo is taken as an NFW profile. While both models fit the data
equally well, the former (i.e. the contracted halo) is more physically
motivated and is also the one whose predictions agree best with other
independent observations. In particular, our contracted halo has
the typical concentration of a ${\sim}10^{12}\Msun$ halo as predicted
by numerical simulations (without imposing any prior on the
concentration), corresponds to a more massive halo than in the pure
NFW case, and also favours a MW stellar mass ${\sim}20\%$ lower than
the NFW case. We show that the two cases can be distinguished using
three diagnostics: i)~the stellar mass of the MW, ii)~the
rotation curve between 1 and $5\kpc$, and iii)~an accurate
determination of the total halo mass.

This paper is structured as follows. In
Section~\ref{sec:MW_components} we describe our model for the various
MW baryonic components. In Section~\ref{sec:halo_contraction} we
characterise how the DM distribution changes in response to the
accumulation of baryons at the halo centre, which we study using
hydrodynamical simulations. Section~\ref{sec:MW_contraction} describes
how much we expect the Galactic DM halo to contract given the
distribution of visible matter in the MW.  Section~\ref{sec:MW_model}
presents our best fit model to the MW rotation curve. The results are
discussed and interpreted in Section~\ref{sec:discussion}. We conclude
with a short summary in Section~\ref{sec:conclusion}.

\vspace{-.3cm}
\section{The MW baryonic components}
\label{sec:MW_components}
The goal of this paper is to infer the mass profile of the MW, and in particular the profile of the DM halo. To do so, we first need
to specify the baryon distribution in the MW, which we model using a
bulge, a thin and a thick stellar disc; an \HI{} disc and a molecular
gas disc; and a diffuse gaseous halo. The first five of this baryonic
components are the same that \citet{McMillan2017} considered, but some
of the parameter values we adopt are different since they correspond
to the best fitting values to the data, as we will describe in
Section~\ref{sec:MW_model}. The mass and profile of the Galactic
gaseous halo (i.e. the circumgalactic medium, hereafter CGM) is
unconstrained; however, both analytical arguments \citep{FW1991} and
  hydrodynamical simulations \cite[e.g][]{Schaye2015}, suggest that
  it contains the majority of the baryonic mass at large distances
  from the Galactic Centre. Section~\ref{subsec:CGM} presents our best
  model for the MW CGM. The MW also has a stellar halo, but its mass
  is insignificant, roughly 3~percent of the total Galactic stellar
  mass \citep{Deason2019b}, and thus we neglect this Galactic
  component.

\vspace{-.3cm}
\subsection{Bulge}
We model the MW bulge using the \citet{McMillan2017} profile
\citep[which is an axisymmetric form of the model proposed
by][]{Bissantz2002} given by, 
\begin{equation}
  \rho_\mathrm{bulge}=\frac{\rho_{0,\mathrm{bulge}}}{(1+r^\prime/r_0)^\alpha}\;
  \textrm{exp}\left[-\left(r^\prime/r_{\mathrm{cut}}\right)^2\right]
  \label{eq:MW_bulge} \;,
\end{equation}
where, $r^\prime$ represents a combination of the cylindrical
coordinates $(R,z)$ (where $R$ is in the plane of the MW disc and $z$
perpendicular to this plane): 
\begin{equation}
  r^\prime = \sqrt{R^2 + (z/q)^2}
  \;.
\end{equation}
The remaining quantities, $\alpha$, $r_0$, $r_{\mathrm{cut}}$ and the
axis ratio, $q$, are model parameters whose values are listed in
Table~\ref{tab:fixed_parameters} and kept fixed for the remainder of
this analysis. The parameter, $\rho_{0,\rm{bulge}}$, denotes the
central stellar density which is allowed to vary according to the
Gaussian prior given in Table~\ref{tab:variable_parameters}. We note
that there is still a large degree of uncertainty regarding the exact
mass and profile of the MW bulge \citep[e.g. see the compilations
of][]{Iocco2015,Bland-Hawthorn2016} and that our data, which cover
only distances beyond $5\kpc$ from the Galactic Centre, are not able to
provide any meaningful constraints on the bulge mass or its radial
profile. Also, for the same reason we do not model the complicated geometry of the stellar distribution at the centre of the MW, i.e. peanut bulge and bar \citep[e.g.][]{Portail2017}, since it has only minor effects on the gravitational field at $R>5\kpc$.

\begin{table}
    \centering
    \caption{ The parameters of the MW components that are kept fixed when fitting our model to observations.
    }
    \begin{tabular}{ @{} l c p{5.1cm} @{} } 
        \hline\hline
        Component & Expression &  Parameters \\
        \hline  \\[-.2cm]
        
        Bulge & Eq. \eqref{eq:MW_bulge} & $r_0=75\pc$, $r_{\rm{cut}}=2.1\kpc$,  $\alpha=1.8$, $q=0.5$ \\ 
        Thin disc & Eq. \eqref{eq:MW_disc} & $z_{\rm{d,\; thin}} = 300\pc$ \\
        Thick disc & Eq. \eqref{eq:MW_disc} & $z_{\rm{d,\; thick}} = 900\pc$ \\
        \HI{} disc & Eq. \eqref{eq:MW_gas_disc} & $z_{\rm d}=85\pc$, $R_{\rm m}=4\kpc$, $R_{\rm d}=7\kpc$, $\Sigma_{0}=53\Msun\pc^{-2}$ \\
        \HII{} disc & Eq. \eqref{eq:MW_gas_disc} & $z_{\rm d}=45\pc$, $R_{\rm m}=12\kpc$, $R_{\rm d}=1.5\kpc$, $\Sigma_{0}=2200\Msun\pc^{-2}$ \\
        CGM & Eq. \eqref{eq:CGM_density} & $A_{\rm CGM}=0.190, \beta_{\rm CGM}=-1.46$ 
        \\
        \hline\hline
    \end{tabular}
    \label{tab:fixed_parameters}
\end{table}

\begin{table*}
    \centering
\caption{ The parameters of the MW components that are varied when
      fitting our model to observations. The columns are as
      follows: parameter description (1) and symbol denoting it (2); 
      units (3); mean and standard deviation of the Gaussian prior
      (4); the MLE and the 68 percentile confidence interval for the model
      with a contracted NFW DM halo (5); and the MLE and the 68
      percentile confidence interval for the model with an
      uncontracted NFW profile for the DM halo (6). For convenience
      and ease of use, the last rows of the table give derived
      quantities, such as bulge, disc and total masses. 
    }

    \renewcommand{\arraystretch}{1.5}
    \begin{tabular}{ p{3.2cm} p{1.7cm}p{2.0cm} ccc } 
        \hline\hline
        Quantity & Symbol & Units & Prior & \multicolumn{2}{c}{Best fitting values} \\[-.2cm]
        & & & & Contracted halo & NFW halo \\
        
        \hline
        bulge density & $\rho_{0,\rm bulge}$ & $\Msun\pc^{-3}$ & $100\pm10$ & $103_{-11}^{+10}$ & $101_{-9}^{+12}$ \\ 
        \rowcolor{Gray}
        thin disc density & $\Sigma_{0,\rm thin}$ & $\Msun\pc^{-2}$ & -- & $731_{-112}^{+91}$ & $1070_{-190}^{+47}$ \\ 
        thick disc density & $\Sigma_{0,\rm thick}$ & $\Msun\pc^{-2}$ & -- & $101_{-65}^{+41}$ &  $113_{-60}^{+50}$ \\ 
        \rowcolor{Gray} 
        thin disc scale length & $R_{\rm thin}$ & $\kpc$ & $2.5\pm0.5$ & $2.63_{-0.12}^{+0.14}$ & $2.43_{-0.07}^{+0.15}$ \\ 
        thick disc scale length & $R_{\rm thick}$ & $\kpc$ & $3.5\pm0.7$ & $3.80_{-.89}^{+0.54}$ & $3.88_{-0.96}^{+0.33}$ \\ 
        \rowcolor{Gray}
        DM mass within $R_{200}$ & $M^{\rm DM}_{200, \rm MW}$ & $10^{12} \Msun$ & -- & $0.97_{-0.19}^{+0.24}$ & $0.82_{-0.18}^{+0.09}$ \\ 
        halo concentration$^\dagger$ & $c^{\rm NFW}_{\rm MW}$ &  & -- & $9.4_{-2.6}^{+1.9}$ & $13.3_{-2.7}^{+3.6}$ \\ 
        
        \hline \\[-.4cm]
        \multicolumn{6}{c}{\bf Derived quantities} \\
        bulge mass & $M_{\star,\rm bulge}$ & $10^{10} \Msun$ &  & $0.94_{-0.10}^{+0.09}$ & $0.92_{-0.08}^{+0.11}$ \\ 
        \rowcolor{Gray}
        thin disc mass & $M_{\star,\rm thin}$ & $10^{10} \Msun$ &  & $3.18_{-0.45}^{+0.30}$ & $3.98_{-0.67}^{+0.26}$ \\ 
        thick disc mass & $M_{\star,\rm thick}$ & $10^{10} \Msun$ &  & $0.92_{-0.12}^{+0.19}$ & $1.07_{-0.19}^{+0.18}$ \\ 
        \rowcolor{Gray}
        total stellar mass & $M_{\star,\rm total}$ & $10^{10} \Msun$ &  & $5.04_{-0.52}^{+0.43}$ & $5.97_{-0.80}^{+0.40}$ \\ 
        \HI{} and molecular gas mass$^\ddagger$ & $M_{\rm{HI + H2}}$ & $10^{10}\Msun$ &  & $1.2$ & $1.2$ \\  
        \rowcolor{Gray}
        CGM mass within $R_{200}$ $^{\parallel}$ & $M_{\rm CGM}$ & $10^{10} \Msun$ &  & $6.4$ & $5.5$ \\ 
        total mass within $R_{200}$ & $M_{\rm 200,MW}^{\rm total}$ & $10^{12} \Msun$ &  & $1.08_{-0.14}^{+0.20}$ & $0.95_{-0.19}^{+0.10}$ \\ 
        \rowcolor{Gray}
        halo scale radius & $R_{s; \ \rm MW}$ & $\kpc$ &  & $23.8_{-6.2}^{+8.1}$ & $14.4_{-3.5}^{+4.5}$ \\ 
        halo radius$^\star$  & $R_{200}$ & $\kpc$ &  & $218_{-18}^{+12}$ & $207_{-15}^{+7}$ \\ 
        \hline\hline
        \\[-.3cm]
     \end{tabular}
     
     \renewcommand{\arraystretch}{1.}
     \begin{tabular}{ @{} p{2.08\columnwidth} @{} }
        $^\dagger$ \changed{The concentration is calculated with respect to $R_{200}$ of the total (DM plus baryons) mass distribution.} For the contracted halo model, the halo concentration corresponds to the value associated to the NFW profile that describes the halo before contraction. \\
        $^\ddagger$ The gas mass has been taken as constant and was
       not varied when fitting our model. We give it here for
       completeness. \\ 
       $^{\parallel}$ The CGM mass is calculated as a fraction of
       $5.8\%$ of the total mass within $R_{200}$ -- see discussion in 
       Section~\ref{subsec:CGM}. \\
      $^\star$ \changed{The halo radius, $R_{200}$, corresponds to the radius of a sphere whose mean enclosed total (DM plus baryons) density is 200 times the critical density.} \\[-.1cm]  
    \end{tabular}
    \label{tab:variable_parameters}
\end{table*}

\subsection{Thin and thick stellar discs}
We model the MW stellar distribution as consisting of two components,
a thin and a thick disc \citep[e.g.][]{Juric2008,Pouliasis2017}, with
each component described by the exponential profile:
\begin{equation}
    \rho_{\mathrm{d}} (R,z) = \frac{\Sigma_{0}}{2z_{\mathrm{d}}} \; \exp\left( -\frac{\mid
      z\mid}{z_{\mathrm{d}}} - \frac{R}{R_{\mathrm{d}}} \right)
    \label{eq:MW_disc} \;,
\end{equation}
where $z_{\rm{d}}$ denotes the disc scale-height, $R_{\rm{d}}$ is the
disc scale-length and $\Sigma_{0}$ is the central surface density. For
the scale-height, we take the values derived by \citet{Juric2008}, who
find that $z_{\rm{d}}=300$ and $900\pc$ for the thin and thick discs
respectively \citep[see also the recent analyses of the Gaia and DES
data:][]{Mateu2018a,Pieres2019}.  We note that the exact value of
$z_{\rm{d}}$ does not significantly affect the inferred MW mass model
-- see e.g. \citet{McMillan2011}. The other two parameters of each
disc model, $R_{\rm{d}}$ and $\Sigma_{0}$, are derived from the data
as we will discuss in Section~\ref{sec:MW_model}. When deriving the
scale-length for both the thin and thick discs, we used the Gaussian
prior given in the fourth column of Table~\ref{tab:variable_parameters},
which is based on the typical scatter in $R_{\rm{d}}$ amongst 
different studies \citep[see the compilation of measurements
in][]{Bland-Hawthorn2016}.

\subsection{\HI{} and molecular discs}
The next two components of the MW are the \HI{} and the molecular gas
distributions, which can account for a significant fraction of the
baryonic mass and, since they have a different geometry from the
stellar component, cannot be easily treated as part of the stellar
disc \citep{Kalberla2008}. Instead, we model these two components
as an exponentially declining disc-like geometry given by
\citep{Kalberla2008}
\begin{equation}
    \rho_{\mathrm{d}} (R,z) = \frac{\Sigma_{0}}{4z_{\mathrm{d}}} \;
    \exp\left( -\frac{R_{\rm m}}{R} - \frac{R}{R_\mathrm{d}} \right) \;
    {\rm sech}^2\left( \frac{z}{2z_{\mathrm{d}}} \right)
    \label{eq:MW_gas_disc} \;,
\end{equation}
where, as in the stellar disc case, $\Sigma_{0}$ denotes the central
surface density, $z_{\rm{d}}$ the scale-height and $R_{\mathrm{d}}$
the scale-length of the disc. This disc has a inner hole whose size is
controlled by the scale-length, $R_{{\rm m}}$. In general, the mass
and geometry of the MW gas distribution are still uncertain
\citep[e.g. see discussions in][]{Kalberla2008,Heyer2015}; however,
they are reasonably well known at the Sun's position. We take the
\HI{} and molecular gas parameters from \citet{McMillan2017}
determined by matching the two gas discs to observational constraints
around the Sun's position. For completeness, we give the values of
these parameters in Table~\ref{tab:fixed_parameters}. They
correspond to an \HI{} mass of $1.1\times10^{10}\Msun$ and a molecular
gas mass of $10$ percent of the \HI{} mass.

\subsection{Circumgalactic medium}
\label{subsec:CGM}

Galaxies are surrounded by an extended gaseous corona, the CGM, which
consists mostly of hot, diffuse gas but also contains denser, colder
clouds, some moving at high velocity. Due to its diffuse nature, the
CGM is difficult to characterise in detail, and even more so in the
case of our own galaxy where much of the X-ray emission from the hot
gas is absorbed by neutral hydrogen in the disc \citep[for details see
the review by][]{Tumlinson2017}. However, the CGM can contain a large
fraction of the baryonic mass within the diffuse halo and thus needs to
be included when modelling the mass profile of the MW. Note that the
CGM mostly contributes to the baryonic mass profile at large
distances, $r\gtrsim 100\kpc$, from the Galactic Centre, while in the
inner part most of the baryons are found in the disc. For our
study, including the CGM does not significantly alter the inferred DM
halo mass or concentration since these are mostly determined by the
stellar circular velocity curve -- see discussion in
Section~\ref{sec:MW_model}. However, the CGM does affect, at the
${\sim}5$ percent level, the total mass within the radius,
$R_{200}$, as well as the escape velocity at the Sun's position, which
is determined by the total mass profile out to a distance of
$2R_{200}$ \citep[see][]{Deason2019}.

Observationally, the total mass and density profile of the CGM in
MW-mass galaxies are poorly determined and this is likely to remain so
for years to come \citep[e.g.][]{Tumlinson2017}. However, we can use
hydrodynamical simulations to place constraints on the Galactic CGM. For this, we
have measured in the three simulations described in
Section~\ref{subsec:simulations}, \auriga{}, \apostle{} and \eagle{},
the baryonic profile at distances, $r>0.15R_{200}$, which, for the MW,
would correspond to $r\gtrsim30\kpc$. We find significant halo-to-halo
scatter, which is indicative of the diversity of CGM distributions
around MW-mass galaxies \citep{Hani2019,Davies2019a}, but the median
distribution shows good agreement between the three simulations. In
particular, we find that the CGM mass within the halo radius,
$R_{200}$, represents $5.8\pm1.5\%$ of the total mass fraction, while
within $2R_{200}$ the CGM mass fraction increases to $11.5\pm2.5\%$ of
the total mass (the errors correspond to the 68\% confidence interval
and are due to halo-to-halo scatter).  In terms of the cosmic mean
baryon fraction, $\fbar{}=15.7\%$ for a \citet{planck2014} cosmology,
the CGM corresponds to $37$ and $73\%$ of the baryon budget expected
within $R_{200}$ and $2R_{200}$ respectively if the baryons followed
the DM distribution. 

We have assumed that the CGM radial density profile can be expressed
as a power law of distance, i.e.
$\rho_{\rm CGM} \sim r^{\beta_{\rm CGM}}$, and then, taking the CGM
mass fractions within $R_{200}$ and $2R_{200}$ to be $5.8$ and
$11.5\%$ respectively, we have estimated the power-law exponent as
well as the overall density normalisation. The resulting CGM density
is given by:
\begin{equation}
    \rho_{\rm CGM} = 200\rho_{\rm crit} \ A_{\rm CGM} \ \fbar{} \left( \frac{r}{R_{200}} \right)^{\beta_{\rm CGM}}
    \label{eq:CGM_density} \;,
\end{equation}
where $\rho_{\rm crit}$ is the critical density of the Universe,
$A_{\rm CGM}=0.190$ is a normalization factor, and $\beta_{\rm  CGM}=-1.46$ is the index of the power law. Then, the enclosed CGM
mass within radius, $r$ is, given by: 
\begin{align}
    M_{\rm CGM}(<r) = \frac{3A_{\rm CGM}}{\beta_{\rm CGM}+3} \ \fbar{} \ M_{200}^{\rm tot} 
    \left( \frac{r}{R_{200}} \right)^{\beta_{\rm CGM}+3}
    \label{eq:CGM_enclosed_mass} \;,
\end{align}
where $M_{200}^{\rm tot}$ is the total mass within the halo radius
$R_{200}$. For example, if the MW total mass is
$1.0\times10^{12}\Msun$, then the CGM mass within the halo radius is
$5.9\times10^{10}\Msun$, which is almost equal to the inner baryonic
mass, that is  the sum of the stellar components and the \HI{} and \HII{} gas discs.

\section{DM halo response to the central galaxy}
\label{sec:halo_contraction}

We now summarise the details of the three galaxy formation
simulations, \auriga{}, \apostle{} and \eagle{}, which we use to
characterise the changes in the structure of DM haloes that result
from the assembly of a galaxy at their centre. In
Section~\ref{subsec:sim_halo_contraction} we compare each host
halo in the hydrodynamics run with its counterpart in the DM-only
(DMO) run. The goal is to find a parametric expression for the halo
radial density profile given a distribution of baryons and then test
how well it reproduces the contraction of individual DM haloes.

\vspace{-.2cm}
\subsection{Simulations}
\label{subsec:simulations}
The \auriga{} and \eagleS{} simulations assume the
\citet{planck2014} cosmological parameters: $\Omega_m =0.307$,
$\Omega_b=0.048$, $\Omega_\Lambda=0.693$ and
$H_0=100~h \kms\Mpc^{-1}$, with $h=0.6777$. The \apostle{} project
assumes the WMAP7 cosmology \citep{Komatsu2011}, with parameters:
$\Omega_m =0.272$, $\Omega_b=0.045$, $\Omega_\Lambda=0.728$ and
$h=0.704$.  In all the simulations, haloes are identified using the
FOF algorithm \citep{Davis1985} with a linking length $0.2$ times the
mean particle separation and further split into gravitationally bound
substructures using the SUBFIND code \citep{Springel2001}. We study
only central galaxies, i.e. the most massive SUBFIND object associated
with an FOF halo, whose centre is taken to be their most
gravitationally bound particle. The haloes are characterised by the
radius, $R_{200}$, of a sphere whose mean enclosed density is 200
times the critical density, and by the mass, $M_{200}$, contained
within this radius.

\subsubsection{\auriga{}}
\auriga{} is a suite of high-resolution magneto-hydrodynamical simulations of
MW-mass haloes ran with the AREPO code \citep{Springel2010}. The suite
consist of 40 haloes, 30 of which have mass,
$M_{200}\in[1,2]\times10^{12}\Msun$, and were first introduced in
\citet{Grand2017}, plus 10 additional lower mass haloes, with
$M_{200}$ masses just below ${\sim}10^{12}\Msun$ \citep{Grand2019b}. The \auriga{}
systems are zoom-in resimulations of MW-mass haloes selected from the
\eagleS{} $100^3\Mpc^3$ periodic cube simulation \citep{Schaye2015}
that are relatively isolated at $z=0$, that is have no objects more
massive than half their halo mass within a distance of
$1.37\Mpc$. See \citealt{Grand2017} for more details, as well
as for illustrations and properties of the central galaxies in the
\auriga{} haloes.

The \auriga{} simulations successfully reproduce many properties of
observed central and satellite galaxies, such as the stellar masses
and star formation rates of spirals \citep{Grand2017,Marinacci2017},
the density and kinematics of stellar haloes
\citep{Deason2017,Monachesi2018}, and the luminosity function of MW
satellites \citep{Simpson2018}. Here, we use both resolution levels of
the \auriga{} project: the medium resolution, or level 4, and the
higher resolution, or level 3, simulation-- only 6 systems were
resimulated at this resolution. The level 4 runs have initial gas and
DM particle masses of $5\times10^4\Msun$ and $3\times10^5\Msun$
respectively, and gravitational softening $\epsilon=0.37\kpc$, while
level 3 has a 8 times better mass resolution and  2 times better
spatial resolution.

\subsubsection{\apostle{}}
\apostle{} is a suite of 12 pairs of MW-mass haloes selected to
resemble the Local Group in terms of mass, separation, relative
velocity and local environment \citep{Fattahi2016, Sawala2016b}. They
were selected from a DMO simulation of  a $100^3\Mpc^3$ periodic
cube, known as COLOR \citep{Hellwing2016}, and were resimulated at
three resolution levels. Here we have used the medium resolution runs,
which have an initial gas particle mass of ${\sim}1.2\times10^5\Msun$
and gravitational softening $\epsilon=0.31\kpc$, and the four volumes
(8 haloes in total) simulated at $12$ times higher mass resolution and
$12^{{1}/{3}}$ better spatial resolution. Each \apostle{}
volume contains two galactic-size haloes, corresponding to the MW and
M31, and here we use both haloes of each pair.

The \apostle{} simulations were run with a modified version of the
Gadget 3 code \citep{Springel2005} with the reference \eagleS{} galaxy
formation models \citep{Schaye2015,Crain2015}, which were calibrated
to reproduce the galaxy mass function, galaxy sizes and the relation
between black hole mass and galaxy mass. The \eagleS{} model
reproduces galaxy rotation curves \citep{Schaller2015a}, the bimodal
distribution of star formation rates and the cosmic star formation
history \citep{Furlong2015}, the Hubble sequence of galaxy
morphologies \citep{Trayford2015} and the Tully-Fisher relation over a
wide range of galaxy masses \citep{Ferrero2017}.

\subsubsection{\eagle{}}
We have also used the MW-mass haloes from the L025N0752 box of the \eagleS{} project run with the recal model (labelled as
Recal-L025N0752). We refer to this run as \eagle{} hereafter. This consists of a
cosmological volume simulation in a periodic cube of side-length
$25\Mpc$ with a mass resolution $8$ times better than the fiducial
\eagleS{} simulation. The simulation contains $752^3$ DM particles with mass of $1.2\times10^6\Msun$  and a similar number of baryonic particles with initial mass
$2.3\times10^5\Msun$ respectively, and gravitational softening
$\epsilon=0.35\kpc$ \citep[for more details see][]{Schaye2015}. The
\eagle{} simulation has been run using the same galaxy formation model
as the standard \eagleS{} run, but with recalibrated parameter values
that account for the higher mass resolution of the \eagle{} run. The
\eagle{} galaxies match observed galaxy properties at least to the
same extent (and in some cases better) than the standard \eagleS{}
galaxies \citep[e.g. see][]{Furlong2015,Schaller2015a,Schaye2015}.

The \apostle{} and \eagle{} simulations have a similar implementation
of galaxy formation processes, but use different parameter values,
and thus we expect them to make similar
predictions. There are clear advantages in studying the halo and
galaxies in the two samples, since we can test the robustness of the
results against changes in mass resolution as well as in some of the
parameters describing the subgrid galaxy formation
models. Furthermore, with \eagle{} we can study the effect of galaxy
assembly in a much larger sample of objects than in \apostle{} and
thus better characterise the halo-to-halo variation.

We select from the \eagle{} simulation Galactic mass haloes,
that is halos which, in the DMO version of the simulation, have
mass, $M_{200}\in[0.7,3]\times10^{12}\Msun$, and whose counterpart in
the hydrodynamic simulation is also a main halo. These selection
criteria results in 34 haloes.

\subsection{Sample selection}

For all three simulation suites we make use of the hydrodynamics and DMO
versions. Finding the counterpart of a DMO halo in the hydrodynamic
simulation and viceversa is straightforward since we are only
interested in main haloes, not subhaloes.

Our strategy is to model the MW halo as an NFW profile in the absence
of baryons which is subsequently modified by the Galactic baryonic
distribution. For this we select from the three simulation suites those 
systems whose density profile in the DMO version is well described by
an NFW profile -- this represents most of the haloes in our sample
(78\%). Some haloes are not in equilibrium, typically because of
transient events such as mergers \citep[e.g. see][]{Neto2007};
including such haloes would misrepresent the long-term relation
between the DM distributions in the DMO and hydrodynamics 
simulations so we do not consider them further.

We proceed by fitting an NFW profile \citep{Navarro1996,Navarro1997} given by:
\begin{align}
    \rho(r) & =  \frac{\rho_0 R_s^3}{r(r+R_s)^2} \\
    & \equiv  \frac{M_{200}}{4\pi R_{200}^3}\frac{c^3}{\ln(1+c)- \frac{c}{1+c}} \frac{R_s^3}{r(r + \frac{R_{200}}{c})^2}
    \label{eq:NFW_profile} \; ,
\end{align}
where $\rho_0$ is the characteristic density, $R_s=R_{200}/c$ is the
scale radius and $c$ is the halo concentration. If we know the halo
mass, then the NFW profile is determined by a single parameter, which
can be taken as the concentration (see Equation~\ref{eq:NFW_profile}).  

To find the best fitting NFW profiles, we minimise 
\begin{equation}
    \sigma_{\rm{fit}} = \frac{1}{N-1}\sum_{i=1}^{N} \left( \log\rho_i - \log\rho_{\rm{NFW;\,i}} \right)^2
    \label{eq:NFW_fit_error}\; ,
\end{equation}
where the sum is over all the $N$ radial bins used for the fit. As
argued in previous studies \citep[e.g.][]{Neto2007,Schaller2015a}, we
limit the fits to the radial range $[0.05,1]R_{200}$. We perform
the fitting using a single free parameter: the halo concentration,
$c$. We have also tested two-parameter fits, in which the total mass,
$M_{200}$, is also allowed to vary and found very similar results.  \begin{figure}
        \centering
        \includegraphics[width=\linewidth,angle=0]{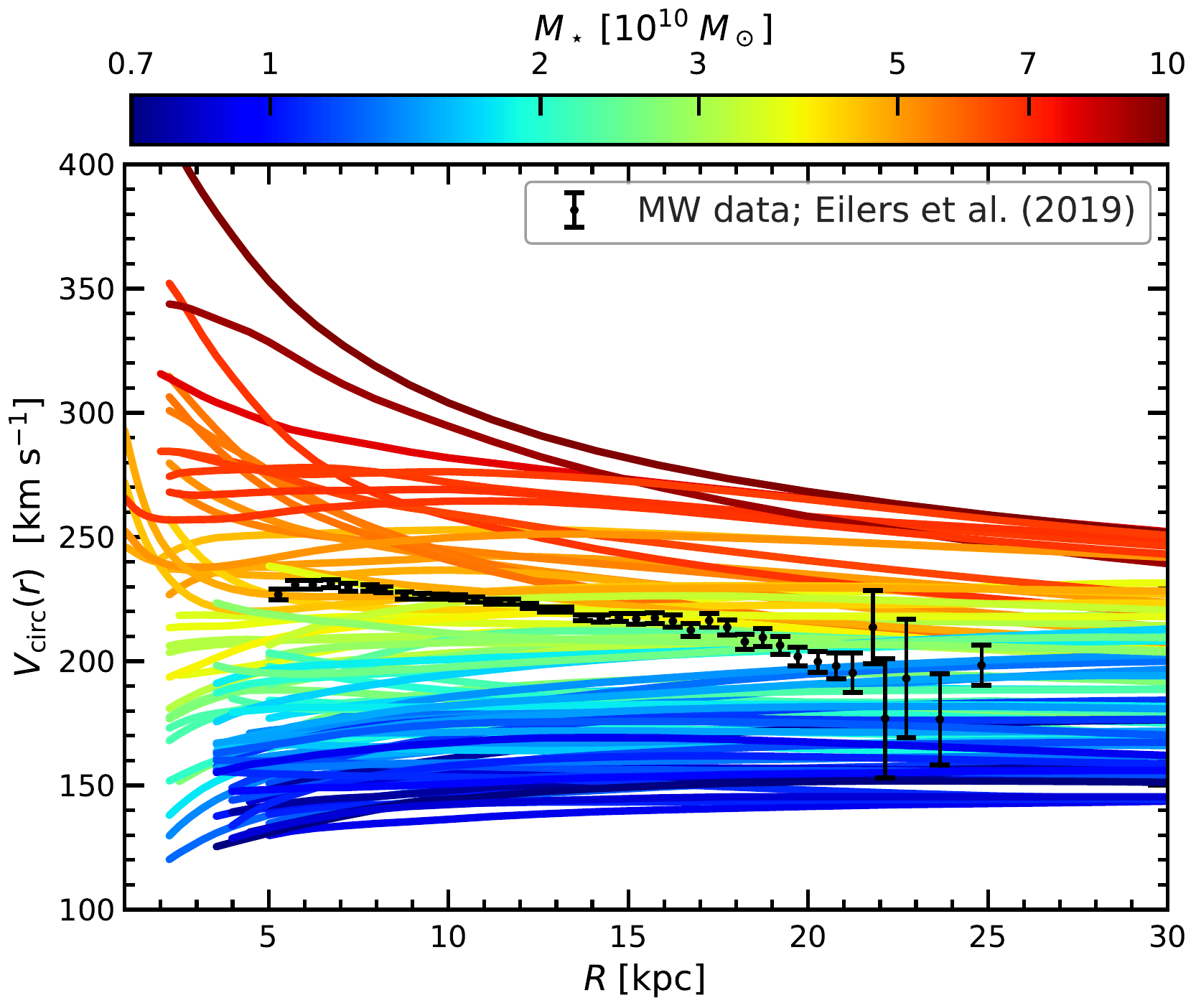}
        \vskip -.2cm
        \caption{Rotation curves for the 87 simulated galaxies used in
          this work. Each line corresponds to one system. The lines
          are coloured according to the stellar mass of the galaxy
          (see legend at the top). The black symbols with error bars
          show the \citet{Eilers2019} determination of the MW rotation
          curve. The error bars correspond to the statistical
          uncertainties associated with the \citeauthor{Eilers2019}
          measurement. For $R>20\kpc$ the MW measurement has large
          (${\sim}10\%$ or higher) systematic uncertainties and thus
          should be interpreted with care.}
          \label{fig:illustration_rotation_curves}
        \vskip -.2cm
\end{figure}

Our final sample is composed of only the haloes whose DMO version is
well described by an NFW profile, which we determine by requiring that
the error in the fit (see Equation~\ref{eq:NFW_fit_error}) be smaller
than $8\times10^{-3}$.  Due to slight stochastic and dynamical
differences between the DMO and full physics simulations, mergers can
take place at slightly different times in matched haloes in the two
simulations. To ensure that we only consider halos in near equilibrium
in the hydrodynamic version we apply the \citet{Neto2007} criterion
to further remove any systems in which the subhalo mass fraction is
higher than 10 per cent. Our final sample consist of 33
medium-resolution and 5 high-resolution \auriga{} haloes, 16
medium-resolution and 6 high-resolution \apostle{} haloes, and 27
\eagle{} haloes.

We account for the limited resolution of the simulations by
considering only regions at $r>2r_{\rm conv}$, where $r_{\rm conv}$ is
the convergence radius from \citet[][see also
\citealt{Power2003}]{Ludlow2019a}. We extend the range to twice the
convergence radius because in hydrodynamics simulations the difference
in the massses of the DM and star particles enhances artificial
two-body scattering \citep[for more details see][]{Ludlow2019b}.

The rotation curves for our sample of 87 simulated galaxies are shown
in Figure~\ref{fig:illustration_rotation_curves}, where they are
compared to the measurement of the MW circular velocity by
\citet{Eilers2019}.
\changed{The rotation curve is measured in the plane of the stellar disc, which is identified with the plane perpendicular to the angular momentum of the stellar distribution within 10\kpc{} from the centre of the galaxy. The velocity is calculated as $ V^2_{\rm circ} = R \ \rm{d} \Phi_{\rm tot}/\rm{d}R$, where $\Phi_{\rm tot}$ is the total gravitational potential and $R$ is the radial distance in the plane of the disc.}
The rotation curve of each simulated galaxy is coloured according to the galaxy stellar mass contained within $10\kpc$ from its centre. Our simulated systems show a diversity of
rotation curves, with maximum values ranging from ${\sim}140$ to
${\sim}300\kms$. The low stellar mass galaxies have low circular
velocities that tend to increase with radius, indicating that their
dynamics are dominated by the DM component. In contrast, the galaxies
with large stellar masses have rotation curves that tend to decrease
with radial distance.

The circular velocities of our simulated galaxies span a range of
values around the measurements for the MW. Some of them are, in fact,
quite close matches to the MW.  In particular, the rotation curves of
simulated galaxies with $M_\star{\sim}4\times10^{10}\Msun$ match the
data well at $R<20\kpc$ (at farther distances the measurements have
large systematic uncertainties that are not shown)
in terms of both absolute value as well as radial gradient. This
stellar mass is in good agreement with estimates for the MW
\citep[e.g.][and Section~\ref{sec:MW_model}]{Bovy2013a,McMillan2017}; 
thus some of our simulated galaxies can be regarded as close
analogues of our galaxy.

\subsection{DM halo profile in the presence of baryons}
\label{subsec:sim_halo_contraction}

\begin{figure}
        \centering
        \includegraphics[width=.98\linewidth,angle=0]{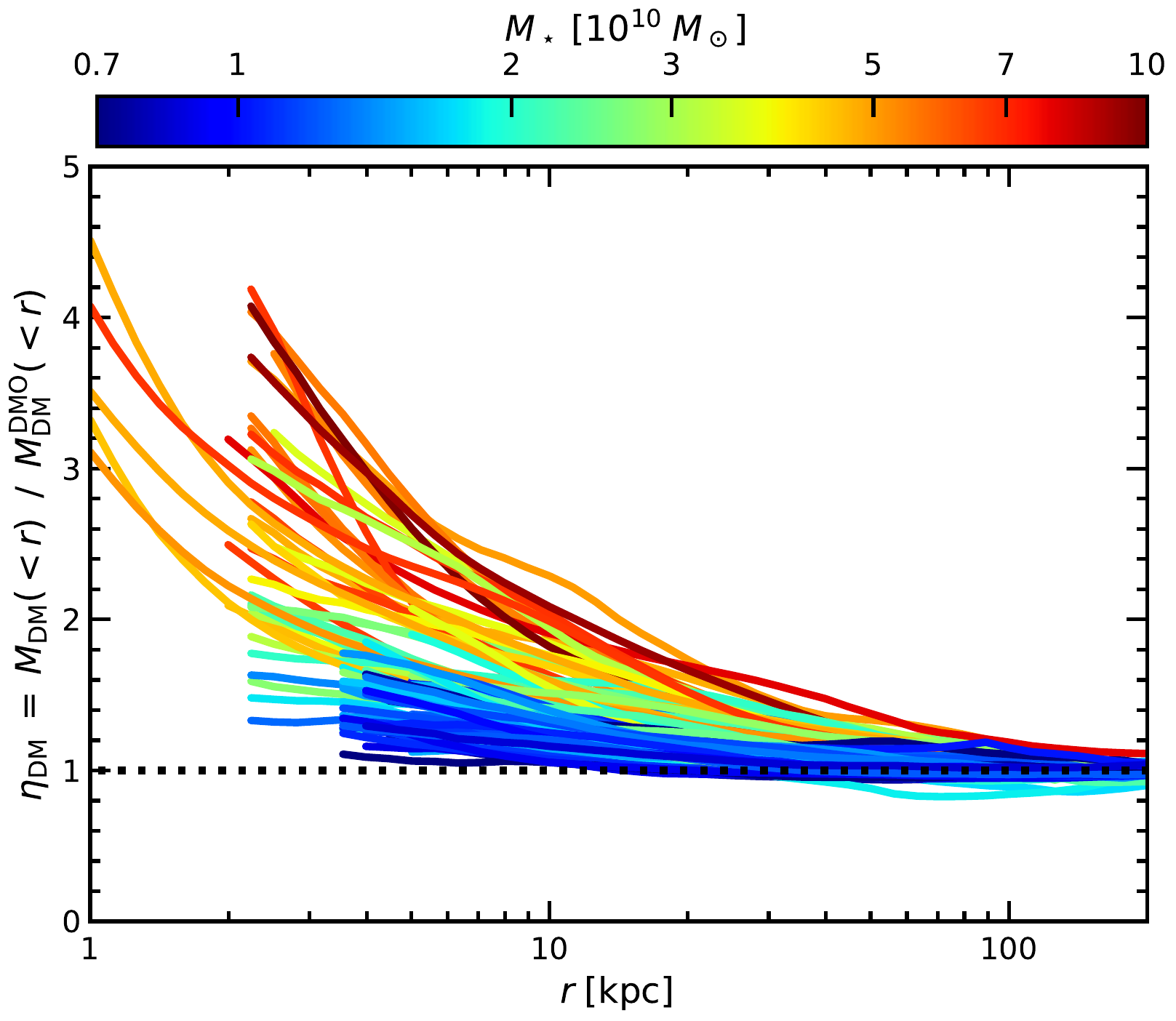}
        \vskip -.2cm
        \caption{ The radial dependence of the ratio, $\eta_{\rm DM}$,
          between the enclosed DM mass in the full physics run,
          $\Mdm(<r)$, and in the DMO only run, $\MdmDMO(<r)$. Each
          line corresponds to a galaxy inside a MW-mass halo from either the \auriga{},
          \apostle{} or \eagle{} hydrodynamical simulations. The lines
          are coloured according to the stellar mass of the central
          galaxy (see colour bar at the top of the panel). We show
          results only for distances larger than that twice the
          \citet{Power2003} radius (see main text). We show results for multiple resolutions, with the highest resolution systems corresponding to the curves that go down to the lowest $r$ values.}
          \label{fig:illustration}
\end{figure}

To study the halo profile in the hydrodynamic simulations, we start by
comparing the enclosed DM mass at different radial distances between
the hydrodynamics run, $\Mdm(<r)$, and the DMO run, $\MdmDMO(<r)$. In
the DMO case all the corresponding mass is associated with a DM
particle but, in reality, each particle should be thought of as
containing a fraction, \fbar{}, of baryons and a fraction $1-\fbar$ of
DM, where $\fbar=\Omega_b/\Omega_m$ is the cosmological baryon
fraction. This implies that the DM mass for the DMO run is given by
$(1-\fbar)\MtotDMO$, where $\MtotDMO$ denotes the total mass in the
DMO simulation.

Figure~\ref{fig:illustration} shows the radial dependence of the
ratio, $\eta_{\rm DM} = \Mdm(<r)/\MdmDMO(<r)$, between the enclosed DM
mass in the hydrodynamics and in the DMO simulations. Each halo in our
three simulation suites is shown as a curve whose colour reflects the
stellar mass, $M_\star$, of the central galaxy. We find that in all
cases the inner $r<10\kpc$ halo is contracted (i.e.
$\eta_{\rm DM}>1$), which implies that the condensation of baryons at
the centre of their haloes leads to an increase in the enclosed DM
mass too. The increase is largest for the most massive central
galaxies. 
Farther from the halo centre we find that some systems still have contracted DM halos, i.e.
$\eta_{\rm{DM}}>1$, while others (especially the ones with low
$M_\star$) have $\eta_{\rm{DM}}<1$, that is less enclosed DM than in
the DMO case. These results are in good agreement with other
hydrodynamics simulations, such as NIHAO \citep{Dutton2016} and
IllustrisTNG \citep{Lovell2018a}, which also show that, on average,
the DM halo is contracted and the amplitude of the contraction
varies among different systems.

\begin{figure}
        \centering
        \includegraphics[width=\linewidth,angle=0]{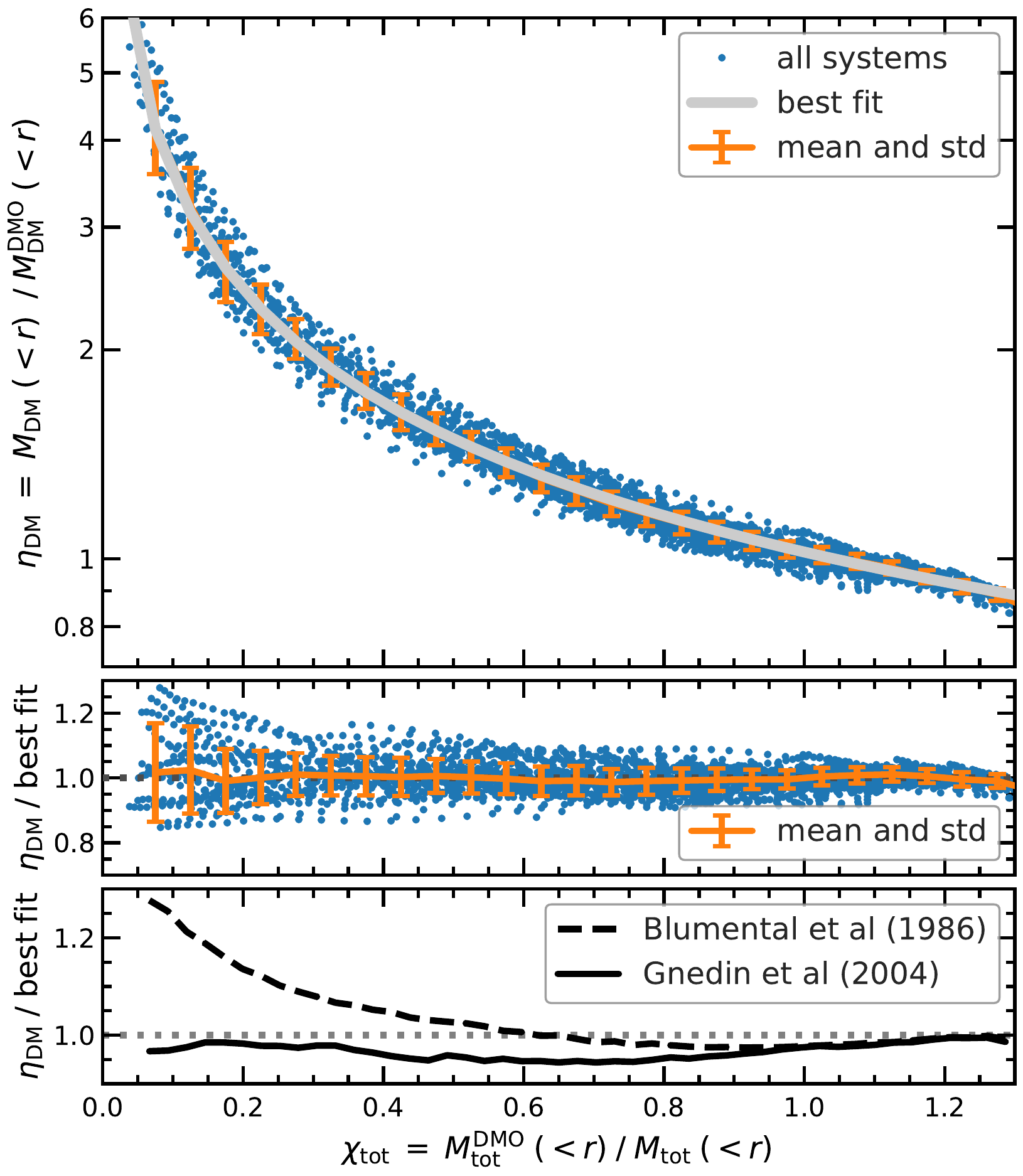}
        \vskip -.2cm
        \caption{ The DM halo response to the assembly of its central
          galaxy. \textit{Top panel}: the ratio of the enclosed DM
          mass, $\eta_{\rm DM} = \Mdm/\MdmDMO$, between the baryonic
          and DMO runs as a function of the ratio,
          $\chi_{\rm tot}=\MtotDMO/\Mtot$, between the total enclosed mass in
          the DMO and the baryonic runs. The DM mass in the DMO run is
          given by $\MdmDMO=(1-\fbar)\MtotDMO$, while the total mass in
          the hydrodynamic run is $\Mtot=\Mdm + \Mbar$. The points
          correspond to $87$ galaxies in three suites of simulations
          whose mass ratios were evaluated at radial distances from
          $1\kpc$ up to $R_{200}$. The thick grey line corresponds to
          the best fitting function described by
          Equation~\ref{eq:DM_increase_fit}. This sits on top of the
          running mean, which is show by the orange
          line. \textit{Centre panel:} the ratio between the
          individual points and the best fit function. The orange line
          with error bars shows the running mean and $68$ percentiles
          of the distribution. \textit{Bottom panel:} comparison
          with the mean $\eta_{\rm DM}$ predicted by the
          \citet{Blumenthal1986} (dashed line) and \citet{Gnedin2004}
          (solid line) approximations to an adiabatically contracted
          halo. 
          }
          \vskip -.2cm
          \label{fig:DM_fraction}
\end{figure}

The response of the DM halo to the assembly of its galaxy can be
predicted to good approximation using the adiabatic contraction method
in which the DM distribution is assumed to have the same action
integrals in the hydrodynamic run as in the DMO case \citep[][the
latter study has explicitly tested this prediction with the \auriga{}
galaxies] {Sellwood2005,Callingham2019b}. However, as we 
discussed in the Introduction, this is a rather involved and
needlessly complicated process. Other simpler adiabatic contraction
approximations, such as those used by \citet{Blumenthal1986} and
\citet{Gnedin2004}, tend systematically to under- or overpredict the
halo contraction
\citep[e.g.][]{Abadi2010,Duffy2010,Pedrosa2010,Dutton2016,Artale2019}. 
In the following, we provide a new description of how the DM halo responds to galaxy formation processes, that combines the simplicity of approximate methods with the accuracy of more involved ones.

We have studied the change in the DM profile as a function of the change in gravitational potential at fixed $r$ between the DMO and the hydrodynamic simulations, which is given by $\chi_{\rm tot}=\MtotDMO(<r)/\Mtot(<r)$ (the mass with a DMO prefix is for the DMO only runs and the one without a prefix is for the hydrodynamics runs). 
We have found that the ratio of the enclosed DM mass, $\eta_{\rm DM} =
\Mdm(<r)/\MdmDMO(<r)$, at a given distance, $r$, is highly correlated
with $\chi_{\rm tot}$. This relation is shown in
Figure~\ref{fig:DM_fraction}, where each data point corresponds to the
pair of $(\chi_{\rm tot},\eta_{\rm DM})$ values for each galaxy
measured at different distances from the centre. The tight correlation
of the $(\chi_{\rm tot},\eta_{\rm DM})$ values is especially
surprising since the same $\eta_{\rm DM}$ value can correspond to
measurements at very different physical radii, depending on the
stellar mass of a galaxy. 
Figure~\ref{fig:DM_fraction} includes galaxies from the three simulation suites studied here: \eagle{}, and both the medium and high resolution runs of \auriga{} and \apostle{}. Although not shown, we have compared
the various resolutions and found very good agreement amongst them
indicating that our results do not depend on numerical resolution. 
\changed{We have also compared disc and spheroidal galaxies and did not find any statistically significant difference between the two morphologies.}

The mean trend between $\chi_{\rm tot}$ and $\eta_{\rm DM}$ (see
solid orange line in Figure~\ref{fig:DM_fraction}) is well captured by
the power-law:
\begin{equation}
    \eta_{\rm DM} = A \ \chi_{\rm tot}^B
    \label{eq:DM_increase_fit} \;,
\end{equation}
with best-fit parameters, $A=1.023\pm0.001$ and  $B=-0.540\pm0.002$.
The best fit function is show by
the grey line in the top panel of Figure~\ref{fig:DM_fraction} which
sits exactly on the median trend (i.e. the orange line). To better
appreciate the quality of the fit, the centre panel of the figure
shows the ratio between the individual data points and the best-fit
function. 
We emphasise that Equation~\ref{eq:DM_increase_fit} has been found for galactic mass halos, i.e. with masses $M_{200}\sim1\times10^{12}\Msun$, and remains to be checked if the same expression can describe the contraction of halos outside this mass range. 

The bottom panel of Figure~\ref{fig:DM_fraction} compares our measured
relation between $\chi_{\rm tot}$ and $\eta_{\rm DM}$ with the
predictions of two widely employed approximations for adiabatic
contraction. We find that both the \citet{Blumenthal1986} and
\citet{Gnedin2004} methods underestimate the DM halo contraction at
high $\chi_{\rm{tot}}$ values, while for $\chi_{\rm{tot}}<0.5$ the
results are mixed. In particular, for $\chi_{\rm{tot}}>0.2$ both
methods are accurate at the 5 per cent level, and while this level of
agreement might seem good, the systematic offset is actually larger
than the typical standard deviation in the individual data points
(see vertical error bars in the middle panel). Note that a 5 percent
error in the relation between $\chi_{\rm tot}$ and $\eta_{\rm DM}$
translates into roughly a $10$ percent error in the determination of
\Mdm. 

\begin{figure}
        \centering
        \includegraphics[width=.97\linewidth,angle=0]{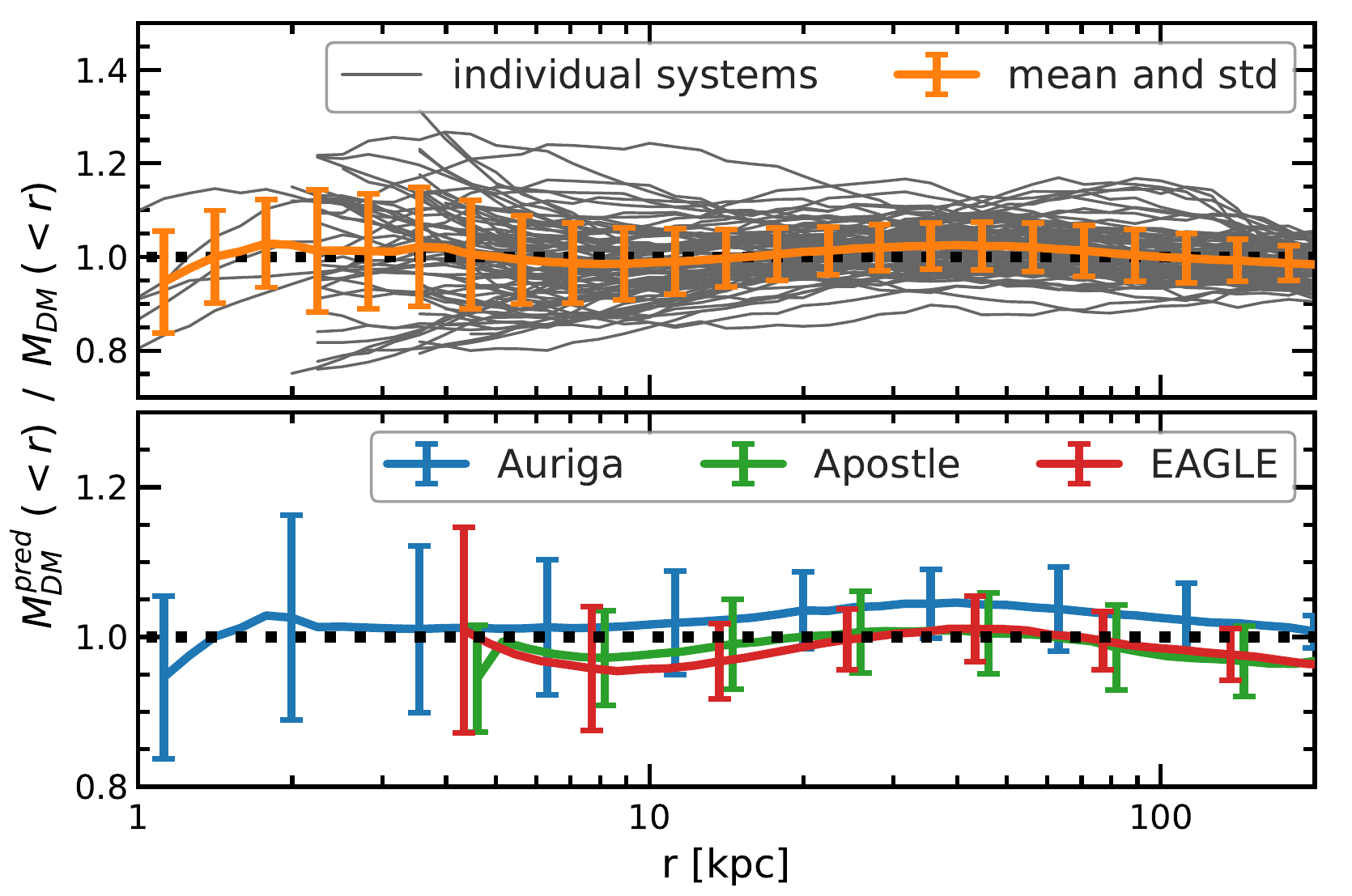}
        \vskip -.2cm
        \caption{ Test of the extent to which our method can recover
          the contracted DM distribution as a function of radial
          distance. The vertical axis shows the ratio between the
          predicted enclosed DM mass, $\MdmPred(<r)$, and the value
          measured in the hydrodynamics simulation, $\Mdm(<r)$. The
          predicted DM mass is calculated from an NFW fit to the
          corresponding halo in the DMO run. The top panel shows
          individual galaxies (grey lines) as well as the mean and the
          $68$ percentiles of the distribution (thick orange
          line). The bottom panel compares the mean and the $68$
          percentiles for galaxies in each of the three simulation suites
          used here: \auriga{} (blue line), \apostle{} (green line)
          and \eagle{} (red line). Our method for inferring the DM
          halo contraction is unbiased and works equally well for all
          three simulations. The halo-to-halo scatter grows from $5\%$
          at $r=100\kpc$, to $7\%$ at $r=10\kpc$ and reaches $13\%$ at
          $r=2\kpc$. 
        }
        \vskip -.2cm
          \label{fig:method_check}
\end{figure}

Equation~\ref{eq:DM_increase_fit} represents a non-linear
deterministic relation between the enclosed mass ratios,
$\chi_{\rm tot}$, and $\eta_{\rm DM}$, which, in turn, can be
expressed as a relation between $\MdmDMO(<r)$, $\Mdm(<r)$ and
$\Mbar(<r)$. Thus, given any two radial mass profiles, we can solve
for the third. For example, we can predict the DM mass profile in the
full physics simulation, $\Mdm(<r)$, given the DM profile in the
absence of baryons and the final baryonic profile. This is exactly
what we are interested in doing here, since we know that $\MdmDMO(<r)$ is
well described by an NFW profile while $\Mbar(<r)$ can be inferred
from observations. These two quantities can be combined with
Equation~\ref{eq:DM_increase_fit} to predict $\Mdm(<r)$, whose solution can be approximated as:
\begin{small}
\begin{equation}
   \Mdm(<r) = \MdmDMO(<r) \left[ 0.45 + 0.38 \left( \eta_{\rm bar} + 1.16 \right)^{0.53} \right]
   \label{eq:DM_increase_fit_2} \;.
\end{equation}
\end{small}
The symbol $\eta_{\rm bar} = \Mbar(<r) / \Mbar^{\rm DMO}(<r)$ denotes the ratio between the enclosed baryonic masses in the hydrodynamics and the DMO runs, where $\Mbar^{\rm DMO}=f_{\rm bar}\MtotDMO$.

We finish this section by testing how well
Equation~\ref{eq:DM_increase_fit_2} reproduces the contraction
of the DM halo. For each halo in our sample, we take the $\Mbar(<r)$
profile from the hydrodynamics simulation and take $\MdmDMO(<r)$ as the
best fitting NFW profile to the DM distribution in the DMO run. We
find the predicted DM mass,
$\MdmPred(<r)$, at each $r$, which we then compare against the actual DM mass
distribution measured in the hydrodynamic run, $\Mdm(<r)$. The results
are shown in top panel of Figure~\ref{fig:method_check}. The mean
ratio of predicted and measured DM masses is very close to one at all $r$,
showing that the method is unbiased. Nonetheless, individual haloes
can deviate from the mean prediction since the size of the contraction
is weakly dependent on the assembly history of the system
\citep[e.g.][]{Abadi2010,Artale2019}.  The halo contraction can be best
predicted at large radial distances, where the halo-to-halo variation
is ${\sim}5$ per cent and is dominated by deviations of the DMO
halo from an NFW profile. In the inner parts, individual haloes can
deviate more from our prediction, but still at a reasonably low level,
with a halo-to-halo scatter of $7$ percent at the Sun's position
and $13$ percent at $2\kpc$.

The bottom panel of Figure~\ref{fig:method_check} addresses a crucial
question: do the predictions depend on the galaxy formation model? To
find the answer, we test the accuracy of the method separately for the
\auriga{}, \apostle{} and \eagle{} samples. For each of the three
simulations we show the mean and the dispersion of the ratio between
predicted and measured DM masses as a function of radial distance. We
find very good agreement between \apostle{} and \eagle{}, which was to
be expected since these two simulations use similar galaxy formation
models. We also find good agreement with the \auriga{} sample:
although this is systematically higher, the difference is smaller than
the scatter amongst individual systems.  The response of the DM halo
to the baryonic component depends on the galaxy assembly history
\citep[e.g.][]{Duffy2010,Dutton2016,Artale2019}; the good agreement
between the halo contraction predictions in our three simulations
suites reflects the fact that these simulations have galaxy growth
histories that match observations \citep[see][and discussion
therein]{Furlong2015}.

\vspace{-.2cm}
\section{The contraction of the MW's halo}
\label{sec:MW_contraction}
Shortly, in Section \ref{sec:MW_model}, we will fit the MW rotation curve to infer the baryonic and DM mass profiles of our galaxy. Before doing so, in this section, we present a brief analysis of how important is the DM halo contraction given the baryonic distribution in the MW. Then, in the second part, we study biases and systematic errors
that arise from not accounting for this contraction. In particular, we
compare the MW total mass and DM halo concentration inferred assuming
that the MW halo is well described by an NFW profile --the usual
approach in the literature-- with the values inferred when the DM
halo contraction is taken into account.

To make the results of this section as relevant as possible to our actual Galaxy, we use the best fitting baryonic mass profile for the MW which we infer in Section \ref{sec:MW_model}. This is given in terms of
the MW baryonic components described in Section~\ref{sec:MW_components}
with the parameter values given in Table~\ref{tab:fixed_parameters}
and in the fifth column (labelled ``best fitting values for contracted
halo") of Table~\ref{tab:variable_parameters}.  The enclosed MW baryonic mass as a function of radial
distance is shown by the black line in
Figure~\ref{fig:MW_enclosed_mass}.

\vspace{-.2cm}
\subsection{Galactic halo contraction}
\label{subsec:MW_halo_contraction}

Both the mass and the concentration of the Galactic halo are
uncertain, so we exemplify the DM halo contraction for a range of halo masses
and concentrations. In all cases we assume that, in the absence of
baryons, the MW DM halo is well described by an NFW profile (see the
discussion in the Introduction) which, in the presence of baryons, is
contracted according to the relation introduced in
Section~\ref{subsec:sim_halo_contraction}. 

Figure~\ref{fig:MW_mass_fraction} shows the increase in the enclosed
DM mass due to the presence of baryons at the centre. For example, if
the MW resides in a $1\times10^{12}\Msun$ halo with the average NFW
concentration for this mass, $c^{\rm{NFW}}=9$ (orange line in top
panel), then the baryons lead to an increase in the enclosed mass at
distances $r<50\kpc$. While the increase is largest for small $r$, it
is still significant at larger distances too, as for example the Sun's orbit
encloses twice as much DM, and a $20\kpc$ radius 30 percent more DM
than the uncontracted halo. The shaded region around the orange line
shows the typical halo-to-halo scatter (see
Figure~\ref{fig:method_check}) and illustrates that we can predict,
with a high degree of confidence, that the Galactic halo is
contracted.
\begin{figure}
        \centering
        \includegraphics[width=.95\linewidth,angle=0]{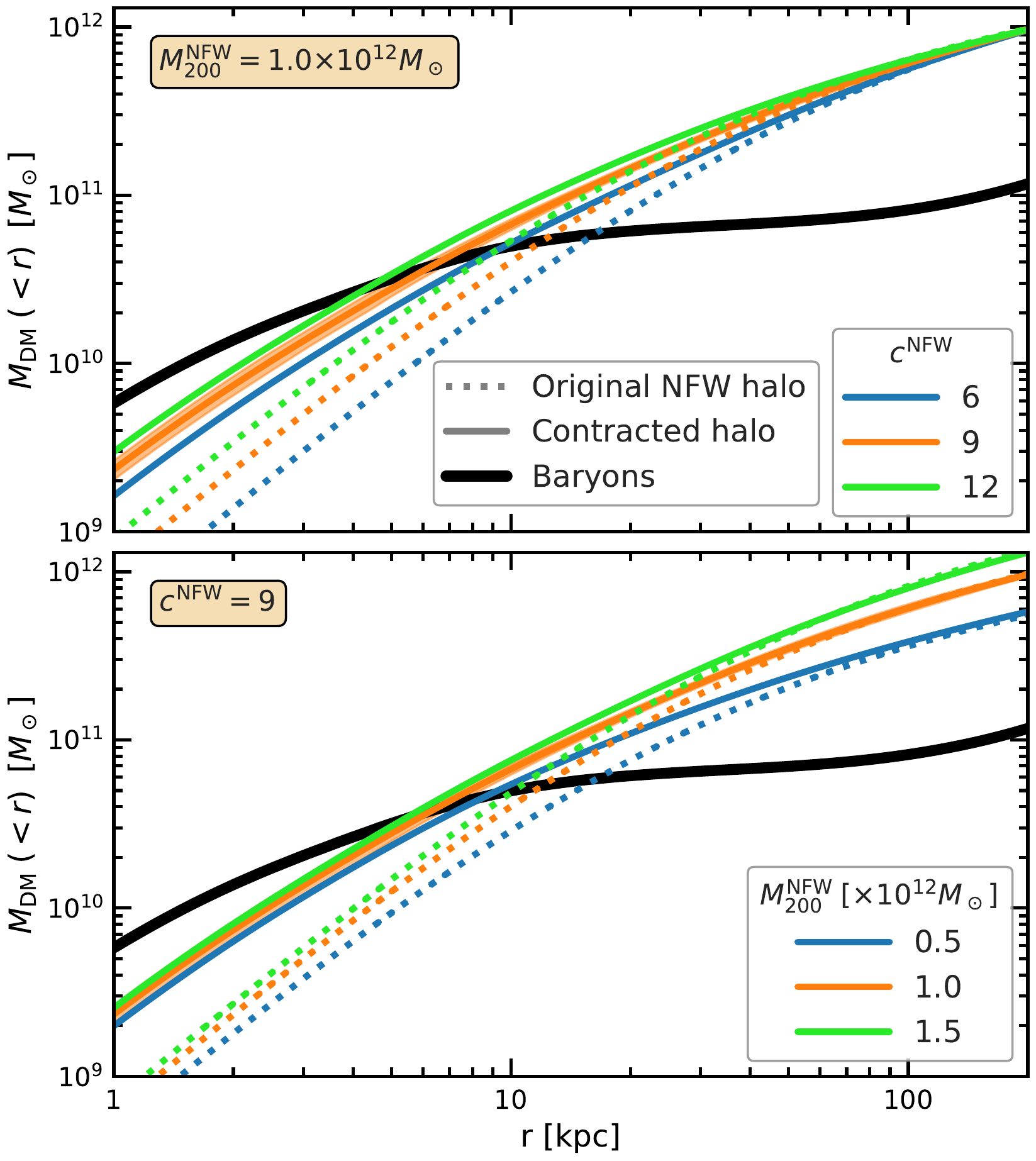}
        \vskip -.2cm
        \caption{The radial enclosed mass profile of NFW haloes
          (dotted lines) and their contracted counterparts (solid lines)
          given the MW baryonic distribution. The solid black line
          shows the Galactic enclosed baryonic mass profile. The top
          panel corresponds to initial NFW haloes of the same mass but
          different concentrations. The bottom panel corresponds to
          haloes with the same concentration but different masses. 
        }
        \label{fig:MW_enclosed_mass}
        \vskip -.2cm  
\end{figure}

At distances, $r>100\kpc$, we notice a small (barely visible) decrease
in the enclosed mass of the contracted halo, which reflects a slight
expansion of the outer halo. This is caused by the fact that at those
distances the enclosed baryonic mass is below the universal baryonic
fraction for the given halo mass and thus the halo experiences the
opposite effect from a contraction: it expands, but only
slightly. Note that while our MW model does include a CGM component,
this is not massive enough to bring up the halo baryonic content to
the cosmic baryon fraction. For example, if the Galactic DM halo mass
is $1.0\times10^{12}\Msun$, then within $R_{200}$ the baryon fraction
is 73\% of the cosmic value. 

\begin{figure}
        \centering
        \includegraphics[width=.95\linewidth,angle=0]{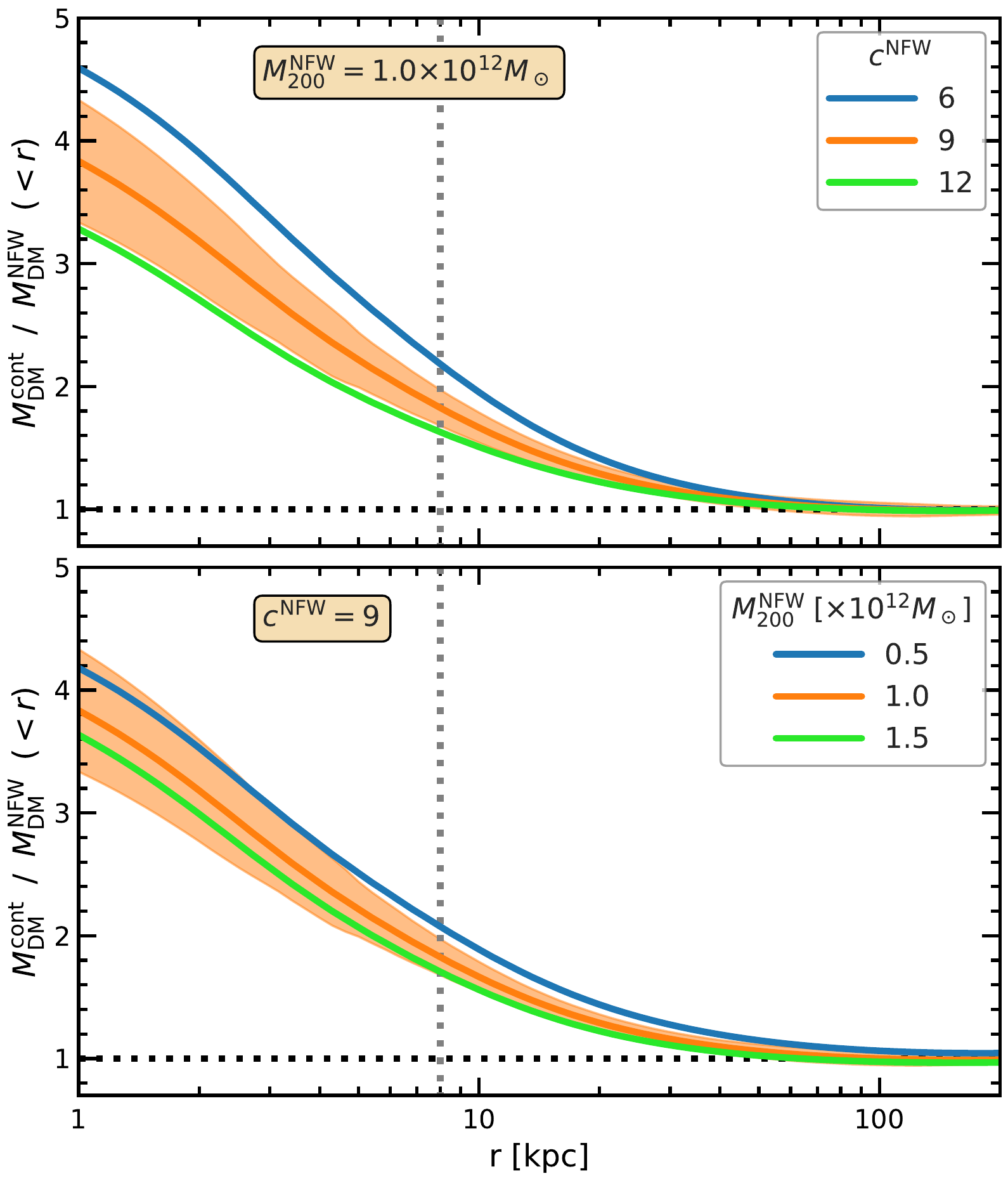}
        \vskip -.2cm
        \caption{ The contraction of the Galactic DM halo for
          different halo masses and concentrations. The Y-axis is the
          ratio of the enclosed DM mass in the contracted halo to that
          in the original NFW halo. In all cases the MW halo, in the
          absence of baryons, is described by an NFW profile of mass,
          $M_{200}$, and concentration, $c^{\rm NFW}$, that is then
          contracted according to the Galactic baryonic
          distribution. The top panel shows haloes with mass,
          $M_{200}=1\times10^{12}\Msun$, and concentrations ranging
          from 5 to 11. The bottom panels shows haloes with
          concentration, $c^{\rm NFW}=9$, and masses ranging from
          $0.5\times$ to $1.5\times$ $10^{12}\Msun$. The orange shaded
          region shows the 68 percentile halo-to-halo scatter in the
          predictions as determined in Figure~\ref{fig:method_check}
          (the scatter is shown only for the orange line). The
          vertical dotted line shows the Sun's position,
          $r_{\odot}=8.2\kpc$.  }
        \label{fig:MW_mass_fraction}
        \vskip -.2cm  
\end{figure}

The top panel of Figure~\ref{fig:MW_mass_fraction} also shows the
contraction of equal mass haloes of different concentrations. The blue
and green curves correspond to concentrations in the absence of
baryons of $c^{\rm{NFW}}=5$ and 11, respectively, which, while falling
in the tails of the $c^{\rm{NFW}}$ distribution, are not very extreme
values. The plot illustrates that the size of the halo contraction
depends sensitively on the halo concentration, with lower
concentration haloes experiencing greater contraction.

The bottom panel of Figure~\ref{fig:MW_mass_fraction} shows that the
size of the contraction also depends on halo mass, but to a lesser
extent than on halo concentration. In this case, the blue and green
curves correspond to DM halo masses of $M_{200}=0.5\times$ and
$1.5\times10^{12}\Msun$, respectively. We find that for the same
baryonic distribution, lower mass haloes contract more.

To understand why the amplitude of the contraction depends on both
halo mass and concentration it is useful to compare the radial profile
of the DM with that of the baryons. This is shown in
Figure~\ref{fig:MW_enclosed_mass} where the thick black line shows the
enclosed baryonic mass, and the various coloured lines show the
enclosed DM mass profile for a range of halo masses and
concentrations. The dotted lines correspond to the original
(i.e. uncontracted) NFW profiles while the solid lines show the
contracted DM distributions. We find that in the inner region, where
baryons dominate, the contraction leads to DM profiles that are much
more similar to one another than to the original NFW
distributions. This implies that the baryons are the main factor that
determines the contracted DM distribution, with the original DM
distribution having a secondary effect. As a result, lower mass or
lower concentration haloes, which have less mass in their inner
regions, must contract more than higher mass or higher concentration
haloes.

\begin{figure}
        \centering
        \includegraphics[width=\linewidth,angle=0]{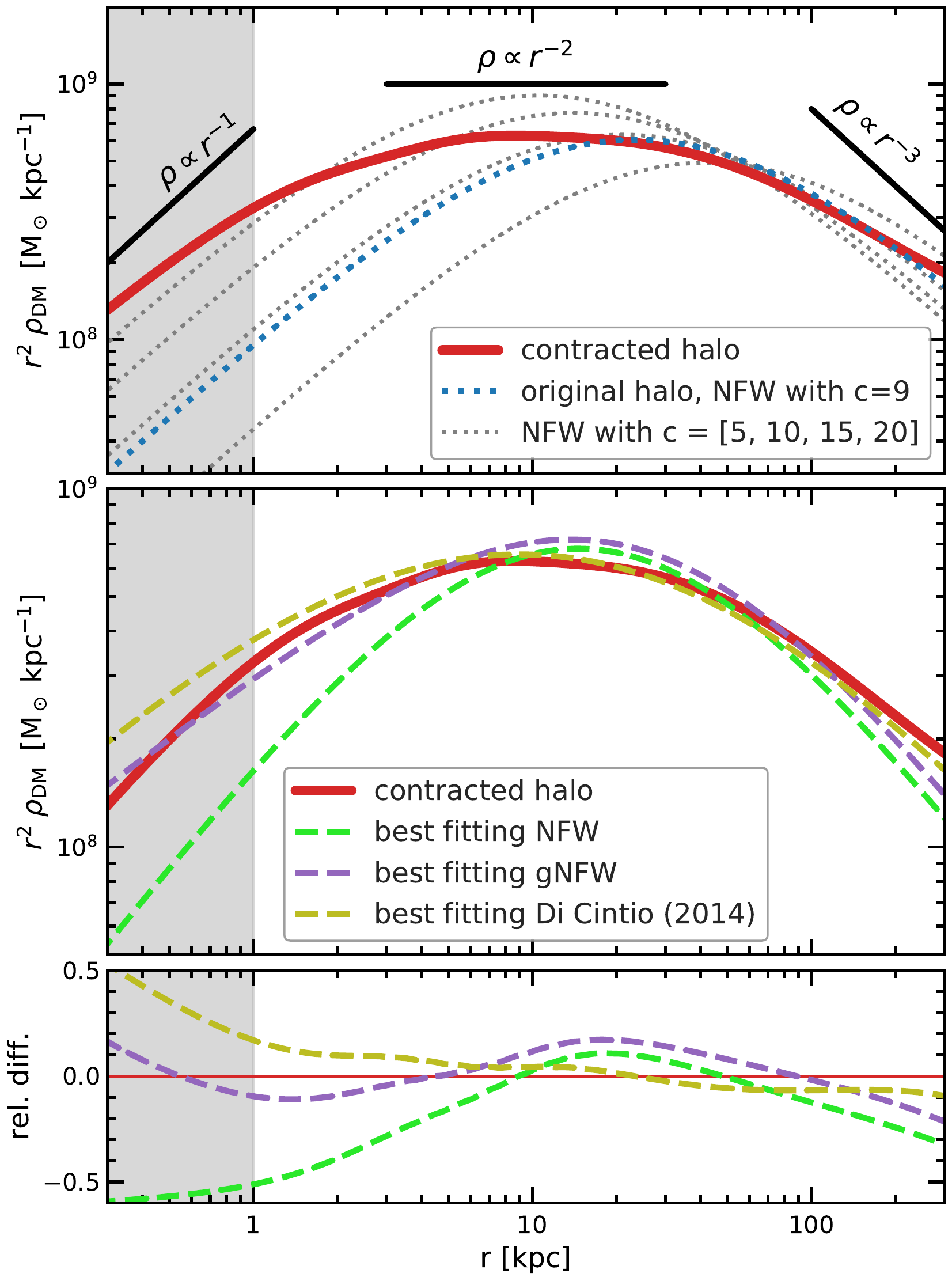}
        \vskip -.2cm
        \caption{ \textit{Top panel:} the density profile of an NFW
          halo (blue dotted curve) of mass,
          $M_{200}=1\times10^{12}\Msun$, and concentration,
          $c^{\rm NFW}=9$, and its contracted counterpart (solid red
          line) given the MW baryonic distribution. This halo profile
          is roughly the same as the best fitting Galactic DM halo
          inferred in Section~\ref{sec:MW_model}. The grey dotted
          lines show NFW profiles for the same halo mass but different
          concentrations.  \textit{Middle panel:} the best fits to the
          contracted Galactic DM halo (solid red line) with an NFW
          (green dashed line), generalised NFW (purple dashed line)
          and \citet[][yellow dashed line]{DiCintio2014}
          profiles. \textit{Bottom panel:} the relative difference, $\rho_{\rm best \ fit} / \rho_{\rm contracted} - 1$,
          between the contracted halo and the three best fitting
          profiles shown in the middle panel.  The grey shaded region
          corresponds to $r<1\kpc$, the regime within which halo
          contraction has been extrapolated to radii smaller than
          those for which we have tested our method. 
          }
          \label{fig:MW_density profile}
\end{figure}

We now investigate if the profile of the contracted halo can be
described by a simple parametric form, such as an NFW profile or more
flexible generalisations. We illustrate this assuming that the MW
galaxy formed in a halo which, in the absence of baryons, is described
by an NFW profile with mass, $M_{200}=1\times10^{12}\Msun$, and
concentration, $c^{\rm NFW}=9$. As we shall see later in
Section~\ref{sec:MW_model}, this halo profile is very close to the
best fitting model for the pre-contracted Galactic halo. The original
NFW halo, as well as its contracted version, are shown in the top
panel of Figure~\ref{fig:MW_density profile} with blue dotted and
red solid lines, respectively. The various gray dotted lines show
NFW profiles for a halo with the same mass but different
concentrations and clearly illustrate that the contracted NFW halo profile
is not of the NFW form.

The middle panel of Figure~\ref{fig:MW_density profile} shows the best
fitting NFW profile, in which both the concentration and the mass are
left as free parameters, to the contracted halo. Since the
contracted halo does not follow an NFW profile, the resulting best
fitting NFW function depends somewhat on the radial range use for the
fit. Here, we fit over the radial range $5\leq r/{\rm kpc} \leq 200$
(the fit is qualitatively similar if we use different reasonable
radial ranges), to obtain the green dashed line in the two bottom
panels. The best fitting NFW form shows large deviations from the
contracted halo profile, ${\sim}20$ percent and even larger,
indicating that an NFW profile is a poor description of a contracted
halo profile. These differences are best illustrated in the bottom
panel of Figure~\ref{fig:MW_density profile}, which shows the relative
difference between the best fitting profiles and the density of the
contracted halo.

We have also tested a more flexible function, the so-called generalised
NFW (gNFW) profile, given by: 
\begin{equation}
    \rho(r) = \frac{\rho_0}{r^\gamma (r+R_s)^{3-\gamma}} 
    \label{eq:gNFW_profile} \; ,
\end{equation}
which, has a third parameter, $\gamma$, in addition to the two
parameters, $R_s$ and $\rho_0$, of the NFW profile. We have fitted the
gNFW profile over the same radial range as the NFW profile to obtain
the purple dashed line shown in the middle and bottom panels of
Figure~\ref{fig:MW_density profile}. The gNFW parametrization does
better at matching the contracted profile in the region $r<5\kpc$,
even though that region was not used in the fit; however, it still
performs poorly at $r>8\kpc$. In particular, the gNFW best fit still
shows a ${\sim}20$ percent deviation from the contracted profile in
the radial range $8\kpc<r<20\kpc$. This is a concern because this
radial range is the sweet-spot between the range for which the MW
rotation curve is least uncertain and the radii at which the DM halo
becomes dominant, so that the data in this intermediate region have
the potential to best constrain the Galactic DM halo.

The inability of an NFW or gNFW function (or other functions such as
an Einasto profile) to describe the contracted profile is a direct
manifestation of the fact that in the radial range,
$5\kpc < r < 30\kpc$, the DM density varies roughly as
$\rho_{\rm{DM}}\propto r^{-2}$ (i.e. $r^2\rho_{\rm{DM}}$ is flat --
see black line in the top panel of Figure~\ref{fig:MW_density
  profile}). The gNFW and Einasto profiles have a range where
$\rho_{\rm{DM}}\propto r^{-2}$, but this is typically limited to a
very narrow interval in $r$, while we predict that the contracted
Galactic DM halo should show this behaviour over a much wider radial
range. More general profiles, such as the \citet{Schaller2015a} or the
\citet{Dekel2017} ones, have more free parameters and potentially can
provide a better match to the contracted halo profile. However, in
practice, their flexibility is also a limitation since the
observational data are not good enough to provide interesting
constraints on the larger number of free parameters (e.g. when fitting
the MW rotation curve, \citealt{Karukes2019} found that the $R_s$ and
$\gamma$ parameters of the gNFW models are highly degenerate). As we
shall discuss in Section~\ref{sec:MW_model}, inferences based on
current MW data already results in $20$ percent uncertainties for
2-parameter DM halo models and these are likely to be even higher for
models with more free parameters.

Some previous works have adopted profiles with several free parameters
and fitted them to the DM density profiles in hydrodynamical
simulations. One example is the study of \citet{DiCintio2014},
who found that a  five parameter profile of the form, 
\begin{equation}
    \rho(r) = \frac{\rho_0}{\left(\frac{r}{R_s}\right)^\gamma \left[1+ \left( \frac{r}{R_s} \right)^\alpha \right]^{(\beta-\gamma)/\alpha}} 
    \label{eq:DiCintio_profile} \; ,
\end{equation}
provides a good description of the DM halo profile in their
hydrodynamic simulations for a wide range of halo masses. In
particular, these authors found that the $\alpha$, $\beta$ and
$\gamma$ parameters in Equation~\ref{eq:DiCintio_profile} depend only
on the stellar-to-halo mass ratio, and thus leaving only two free
parameters, $\rho_0$ and $R_s$. Using the \citet{DiCintio2014}
predicted values for $\alpha$, $\beta$ and $\gamma$, we fitted the
contracted NFW halo distribution in Figure~\ref{fig:MW_density profile}
using Equation~\ref{eq:DiCintio_profile} with two free parameters, 
$\rho_0$ and $R_s$. The resulting best fitting function is shown in
Figure~\ref{fig:MW_density profile} by the yellow dashed line. This
functional form captures the contracted halo profile reasonably well,
with typical errors of 10\% or less. However, these errors are still
larger than the typical uncertainties in the MW rotation curve and
could lead to systematic biases in the inferred halo mass or
concentration.

\subsection{Biases in inferred halo properties}
\label{subsec:MW_bias_halo_properties}

\begin{figure}
        \centering
        \includegraphics[width=\linewidth,angle=0]{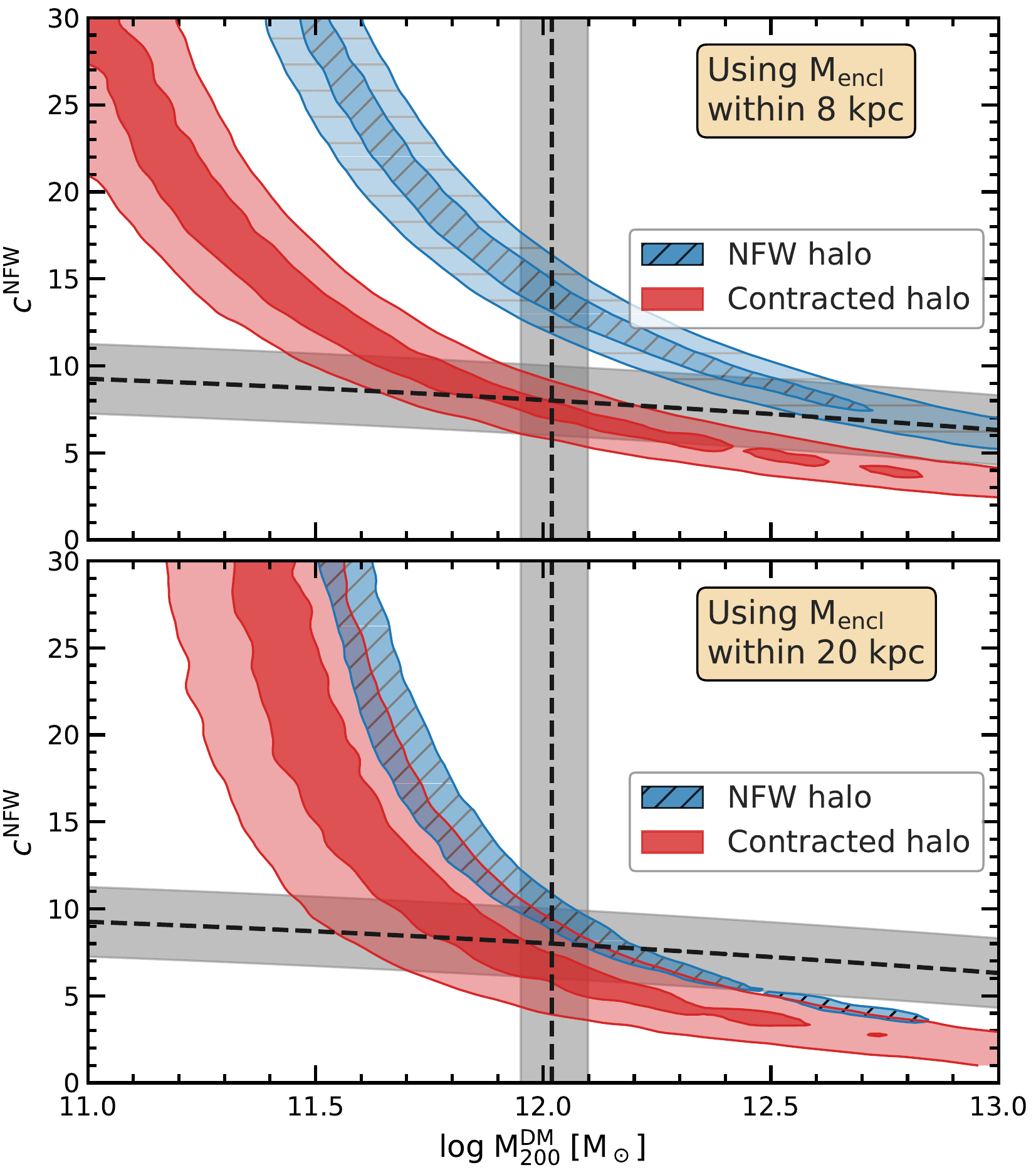}
        \vskip -.2cm
        \caption{ Constraints on the mass and concentration of the MW
          DM halo inferred from the enclosed mass within $8\kpc$ (top
          panel) and within $20\kpc$ (bottom panel). The blue shaded
          region corresponds to modelling the halo as an NFW
          profile. The red shaded region corresponds to modelling the
          halo as an NFW profile that has been contracted by the MW
          baryonic distribution -- in this case the concentration
          corresponds to the original (uncontracted) halo. The dark
          and lighter colours show the 68 and 95 percentile confidence
          regions, respectively. For clarity, for the NFW case in the
          bottom panel, we show only the 68 percentile confidence
          region. The vertical dashed line and the associated grey
          region show the \citet{Callingham2019} MW DM halo mass
          estimate and its 68 percentile confidence region. The
          approximately horizontal dashed line and its associated grey
          region show the median and standard deviation of the
          halo mass--concentration relation \citep{Hellwing2016}.}
          \label{fig:constraint_enclosed_mass}
          \vskip -.2cm
\end{figure}

\begin{figure}
        \centering
        \includegraphics[width=\linewidth,angle=0]{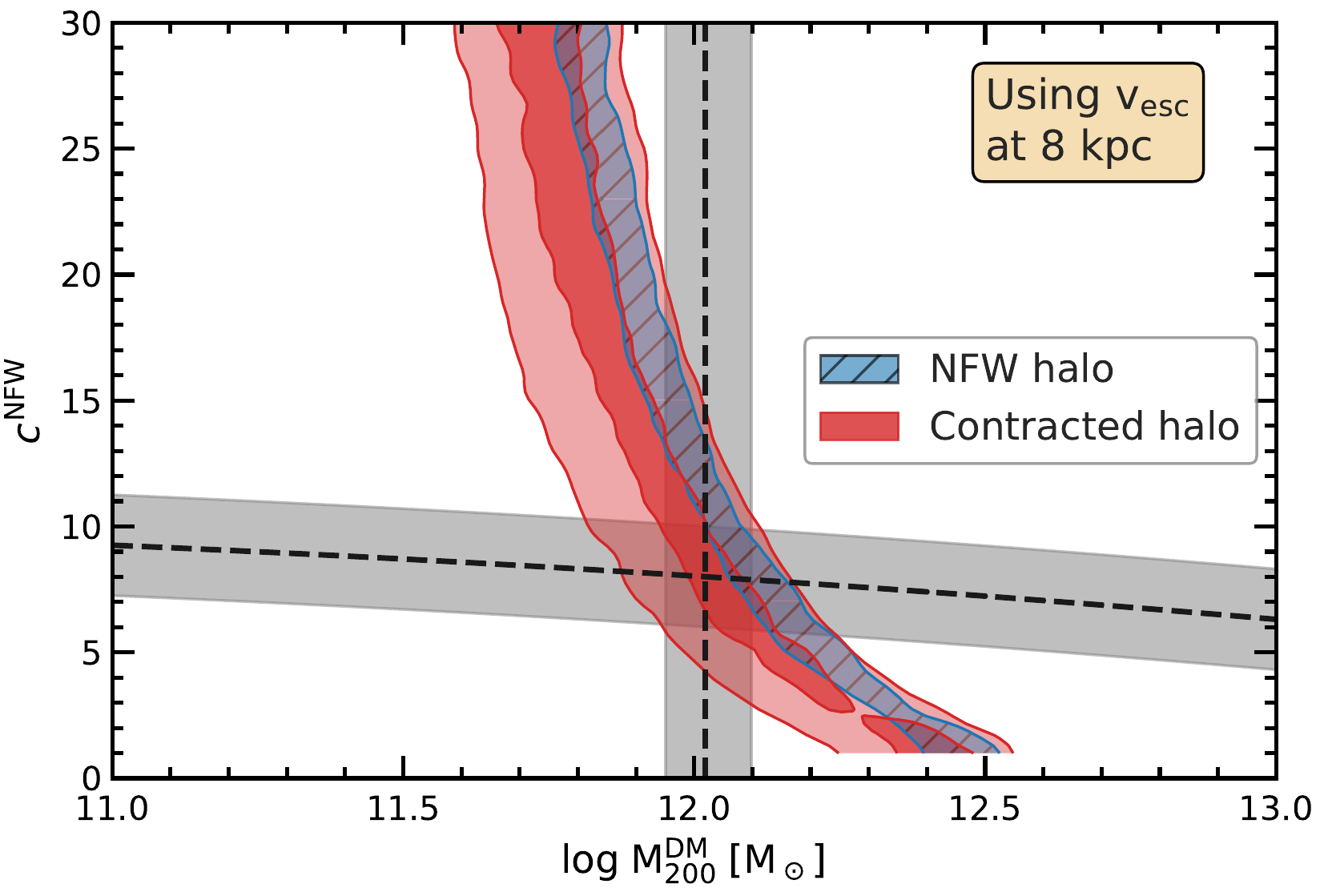}
        \vskip -.2cm
        \caption{ Constraints on the mass and concentration of the MW
          DM halo inferred from the escape velocity measurement of
          \citet{Deason2019}. The blue shaded region shows the 68
          percentile confidence region when modelling the halo as an
          NFW profile. The red shaded regions show the 68 and 95
          percentile contours when taking into account the contraction
          of the Galactic DM halo -- in this case the concentration
          corresponds to the value before applying the baryonic contraction. The
          dashed lines and grey shaded regions are as in
          Figure~\ref{fig:constraint_enclosed_mass}.  }
          \label{fig:constraint_escape_velocity}
\end{figure}

We saw in the previous subsection that the settling of baryons at the
centre of a DM halo causes the halo to contract and, as a result, the
density profile  no longer follows the NFW form. However, many
previous studies have modelled the Galactic halo as an NFW profile,
which raises an important question: what are the biases in the
inferred halo parameters that result when fitting an NFW halo to the
observational data? To answer this question we proceed to study 
how the inferred DM halo mass and concentration differ when the data
are fit with either a contracted NFW halo or an uncontracted NFW profile.

We first infer a DM halo mass and concentration by fitting the
enclosed mass at two different distances from the Galactic Centre, the
Sun's position, $r=8\kpc$, and $r=20\kpc$. We study the enclosed mass
at two radii because the contraction of the halo becomes less
important with increasing distance from the Galactic Centre and thus
systematic differences between a contracted and an NFW halo are
distance dependent.  For simplicity, we assume that there is no uncertainty in the profile of
the baryonic component, and infer the DM halo properties: total mass and concentration (for the contracted halo, the concentration corresponds to the value before contraction). The resulting 68 and 95\% confidence limits
for $M_{200}^{\rm DM}$ and $c^{\rm NFW}$ are shown in
Figure~\ref{fig:constraint_enclosed_mass}. To calculate the enclosed
masses we used the \citet{Eilers2019} circular velocity measurement,
$V_{\rm circ}(r=8\kpc)=(230\pm5)\kms$, and the enclosed total mass
measurement of \citet{Posti2018},
$M^{\rm tot}(<r=20\kpc)=(1.91 \pm 0.18)\times10^{11}\Msun$.

Using a single mass measurement results in a degeneracy between the
inferred halo mass and concentration since different
($M_{200}^{\rm DM},c^{\rm NFW}$) pairs can produce the same enclosed
DM mass, as may be seen from the coloured shaded regions in
Figure~\ref{fig:constraint_enclosed_mass}. More interestingly, the
figure shows that modelling the DM halo as an NFW or a contracted
profile results in very different estimates of the halo mass and
concentration. The difference is especially striking for the estimates
at $r=8\kpc$ (top panel in Figure~\ref{fig:constraint_enclosed_mass}),
where we find that even the 95\% confidence limits for the two models
do not overlap. At larger distances, such as at $r=20\kpc$ shown in
the bottom panel of Figure~\ref{fig:constraint_enclosed_mass}, the
baryons lead to a smaller contraction of the DM halo and the two model
estimates are in closer agreement, but still do not have overlapping
68\% confidence limits.

The ($M_{200}^{\rm DM},c^{\rm NFW}$) confidence regions can be
combined with other measurements or theoretical priors to narrow the
uncertainty regions. For example, the (roughly) horizontal dashed line
and its associated grey shaded region show the halo
mass--concentration relation from DM-only cosmological simulations
(\citealt{Hellwing2016}; this is very similar to other recent
mass--concentration relations, as may be seen from Figure~5 of that
paper). Using the relation as a prior, we can estimate the DM mass of
the Galactic halo. Doing so for the contracted NFW halo model results in a
consistent estimate of $M_{200}^{\rm DM}{\sim}1\times10^{12}\Msun$ for
both $r=8$ and $20\kpc$, which is in good agreement with the recent
estimate by \citet[][vertical dashed line] {Callingham2019}. In
contrast, the NFW halo model prefers a very high DM mass at $r=8\kpc$,
$M_{200}^{\rm DM}{\sim}1\times10^{13}\Msun$, and a much lower mass,
${\sim}1.5\times10^{12}\Msun$, at $r=20\kpc$.

More interesting is to combine the contours in
Figure~\ref{fig:constraint_enclosed_mass} with other DM mass estimates
to infer the concentration of the Galactic DM halo. We illustrate this
by showing the \citet{Callingham2019} DM mass estimate and its
associated 68\% confidence interval, which are shown in the figure as
the vertical dashed line and associated grey shaded region. The
contracted halo model predicts that the MW has an (uncontracted)
concentration, $c^{\rm NFW}{\sim}8$, which is typical of a
$1\times10^{12}\Msun$ $\Lambda$CDM halo -- this can be inferred from
the fact that the vertical and horizontal dashed lines intersect
inside the dark shaded region in both panels in the figure. In
contrast, the inferred concentration for the NFW halo model is very
different for the two radial measurements shown in
Figure~\ref{fig:constraint_enclosed_mass} and is systematically higher
than the theoretical $\Lambda$CDM prediction. Thus, incorrectly
modelling the MW halo using an NFW profile can lead to a large
overestimate of its concentration.

A complementary method for constraining the Galactic DM halo mass is
by measuring the escape velocity, $V_{\rm esc}$, which, despite its name, is not the velocity needed to reach infinite distance with zero speed. \citet{Deason2019} have shown that the escape velocity characterises
the difference in gravitational potential between the position where
$V_{\rm esc}$ is measured and the potential at a distance $2R_{200}$
from the halo centre. The potential depends on the
mass profile of the halo up to $2R_{200}$ and thus modelling the DM
halo as a contracted or an NFW profile can introduce different biases
from those present in enclosed mass measurements. These are studied in
Figure~\ref{fig:constraint_escape_velocity}, where we show the
inferred DM halo properties using the recent measurement of the escape
velocity at the position of the Sun, $V_{\rm esc}=(528\pm25)\kms$, by
\citet{Deason2019}.

Figure~\ref{fig:constraint_escape_velocity} shows that using a NFW
profile instead of a contracted NFW halo also leads to biases in modelling
the escape velocity. Given the current uncertainty in the
$V_{\rm esc}$ measurement, the 68\% confidence regions for the two
models barely overlap; however this will not be the case with for
future large datasets. Compared to
Figure~\ref{fig:constraint_enclosed_mass}, the escape velocity
predictions are less affected by using the incorrect NFW profile since
much of the escape velocity is determined by the mass at large
Galactocentric radii where both the contracted halo and the NFW
profile are very similar. Nonetheless, there are still differences
between these two profiles in the inner region of the halo, which
explains why the incorrect NFW model prefers systematically higher
concentrations than the contracted halo model.

\vspace{ -.2cm}
\section{A total mass model for the MW}
\label{sec:MW_model}
In this section we describe the data and fitting procedure used to
determine the baryonic and DM mass profiles of our galaxy. We perform
the analysis in the same spirit as \citet[][see also
\citealt{Klypin2002a,Weber2010,McMillan2011,Bovy2012a,Kafle2014,McMillan2017}]{Dehnen1998},
that is, we estimate the best fitting MW mass model by varying several
parameters that encode our ignorance about the stellar and DM
distributions of our galaxy. For the DM, we fit two models: a
contracted NFW halo, which is motivated by the predictions of
hydrodynamical simulations (see Section~\ref{sec:halo_contraction}),
and a pure NFW profile, which is one of the most commonly used
profiles in previous studies.

\vspace{ -.2cm}
\subsection{Data}
\label{subsec:MW_model:data}
\begin{figure}
        \centering
        \includegraphics[width=\linewidth,angle=0]{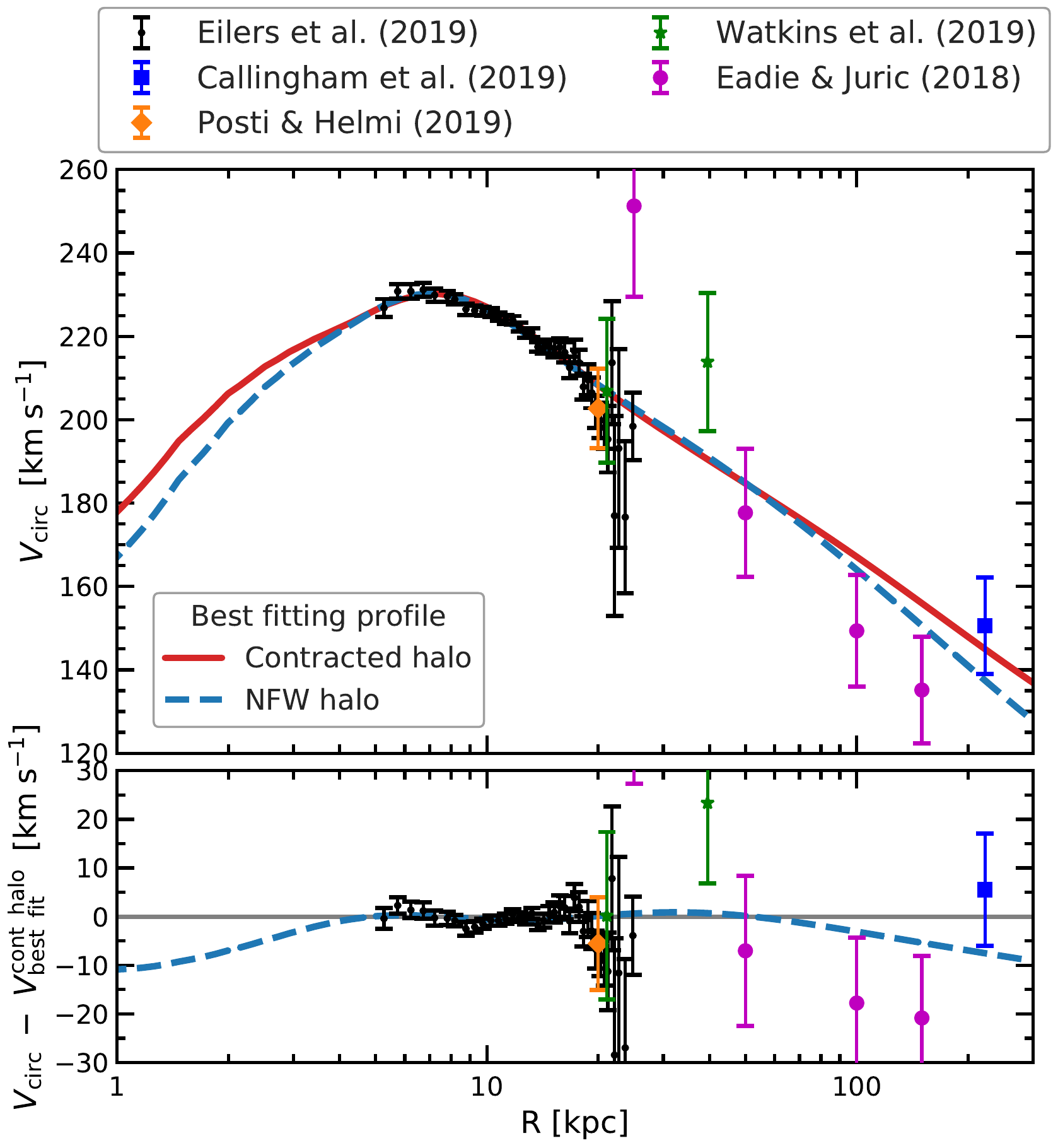}
        \vskip -.2cm
        \caption{ \textit{Top panel:} MW Galactic rotation curve
          (symbols with error bars) as a function of radial
          distance. The solid red line is the best fitting MW mass
          model assuming a contracted DM NFW halo. The dashed blue line
          the best fitting MW mass model assuming no contraction,
          i.e. that the DM halo follows
          an NFW profile. Both models were fitted only to the
          \citet{Eilers2019} and the \citet{Callingham2019} data
          points.  \textit{Bottom panel:} The difference between the
          data and the best fitting contracted halo model. The dashed
          blue line shows the difference between the NFW halo model
          and the contracted halo one. The two models give the same
          rotation curve to within $1\kms$ or less in the range
          $5\kpc<r<60\kpc$.  }
        \label{fig:MW_fit_data}
        
        \vskip .2cm 
        
        \includegraphics[width=\linewidth,angle=0]{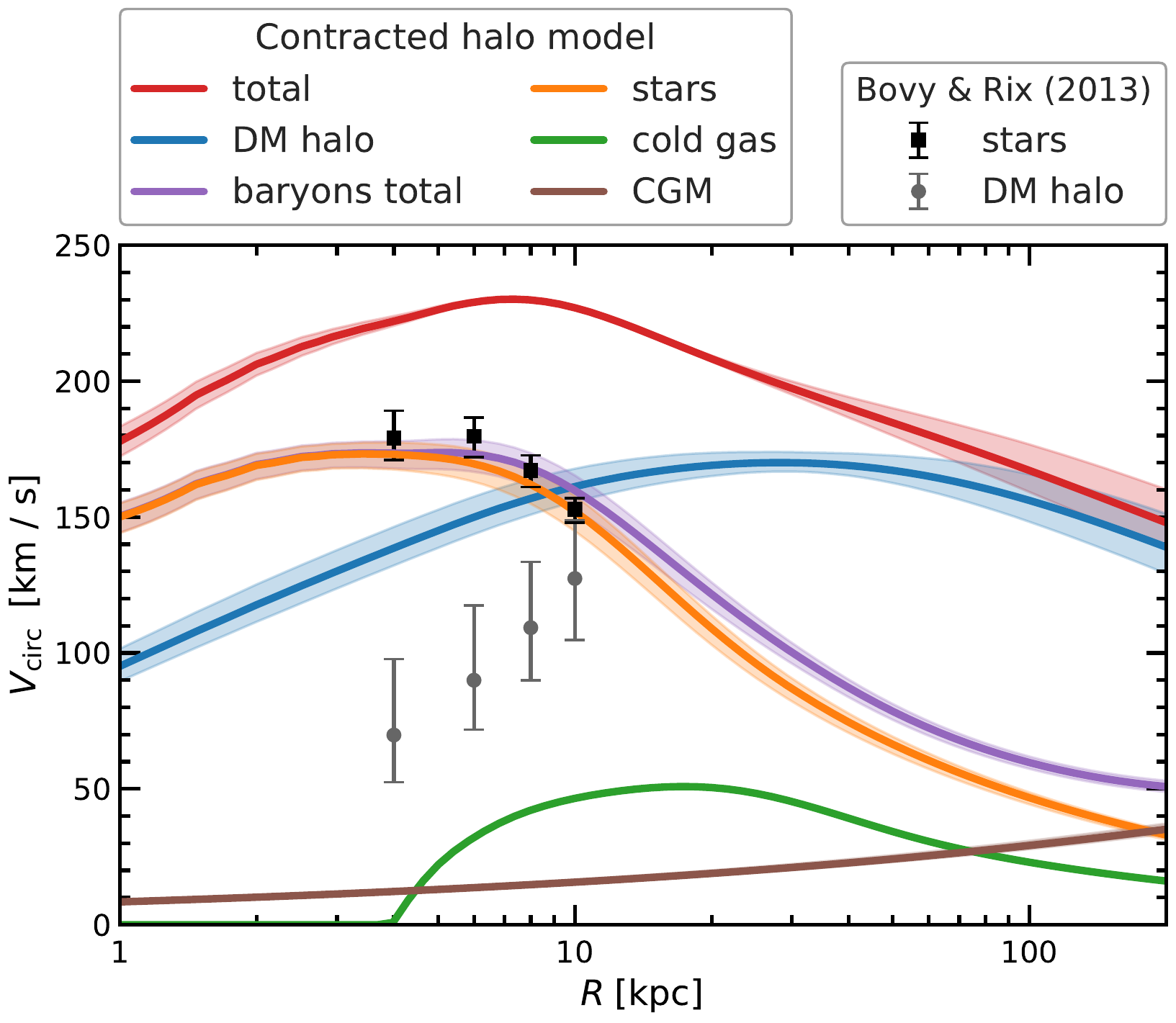}
        \vskip -.2cm
        \caption{ The rotation curve of the best fitting MW contracted
          NFW halo model separated into contributions from individual
          components. The solid lines show the maximum likelihood
          model and the shaded region the 68 percentile confidence
          regions. The symbols with error bars show the
          \citet{Bovy2013a} determination of the stellar disc and DM
          halo of the MW.  }
        \label{fig:MW_best_fit_components}
\end{figure}

The main constraining power of our model comes from the
\citet{Eilers2019} circular velocity data (black data points in
Figure~\ref{fig:MW_fit_data}). These data are inferred from
axisymmetric Jeans modelling of the six-dimensional phase space
distribution of more than 23,000 red giant stars with precise parallax
measurements. The stellar positions and velocities come from a
compilation of \textit{Gaia} DR2 measurements, combined with improved
parallax determinations from APOGEE DR14 spectra and photometric
information from WISE, 2MASS and \textit{Gaia} \citep[for details
see][]{Hogg2018}.
 
The \citeauthor{Eilers2019} rotation curve provides good constraints
in the inner parts of the MW system; however this does not fully break
up the degeneracy between DM halo mass and concentration. To deal with
this, we make use of the total mass estimate of
\citet{Callingham2019},
$M_{200,\rm{MW}}^{\rm total} = (1.17\pm 0.18) \times
10^{12}\Msun$. These authors infer the mass by comparing the observed
energy and angular momentum distribution of the classical MW
satellites with the predictions of hydrodynamical simulations. While
there are many Galactic mass estimates \citep[e.g. see the
compilations in][]{Wang2015a,Callingham2019}, we choose the
\citeauthor{Callingham2019} result since it has several advantages
compared to other studies: i)~the method had been thoroughly tested
with multiple hydrodynamic simulations, ii)~it makes use of the
dynamics of satellites whose extended radial distribution directly
constrains the total mass of the system, and iii)~it makes use of the
latest \textit{Gaia} DR2 proper motion measurements for the classical
dwarfs \citep{GaiaDR22018arXiv180409365G}.

To remove some of the degeneracy between the thin and thick stellar
discs, we impose the prior that the ratio of the thin to thick disc
densities at the Sun's position, which we take as
$R_\odot=8.122\pm0.031\kpc$ \citep{Gravity_Collaboration2018}, is
$0.12\pm0.012$. This value is derived from the analysis of MW disc
stars in the SDSS data by \citet[][]{Juric2008}.

The last measurement we consider is the value of  the vertical force
at $1.1\kpc$ above the plane at the Sun's position, which we take as
\citep{Kuijken1991} : 
\begin{equation}
    K_{z}(R_\odot) = 2\pi G \times (71\pm6) \Msun \pc^{-2}
    \;.
\end{equation}
To implement this constraint, we express it as a function of the local
total surface mass density, $\Sigma$, which is given by
\citep{McKee2015}:
\begin{equation}
    \Sigma = \frac{K_z}{2\pi G} + \Delta \Sigma
    \;,
\end{equation}
where $\Delta\Sigma$ represents a correction term for the fact that
the circular velocity varies with Galactocentric radius and with the
$z$ coordinate above the disc plane. We calculate the $\Delta \Sigma$
term using Eq. (53) from \citet{McKee2015}, combined with the
\citet{Eilers2019} rotation curve to obtain
$\Delta \Sigma = 9 \Msun \pc^{-2}$.

We note that most of the constraining power comes from the
\citet{Eilers2019} circular velocity data. This is due to a
combination of \citeauthor{Eilers2019} having the most data points, 38
in total, and to the fact that most of the measurements are very
precise, with errors below $2\kms$, corresponding to less than 1\%
relative errors. In contrast, the vertical force measurement has an 8\%
relative error, while the total mass estimate has a 15\% relative
error.

\subsection{The fitting procedure}
\label{subsec:MW_model:fitting}

\begin{figure*}
        \centering
        \includegraphics[width=.89\linewidth,angle=0]{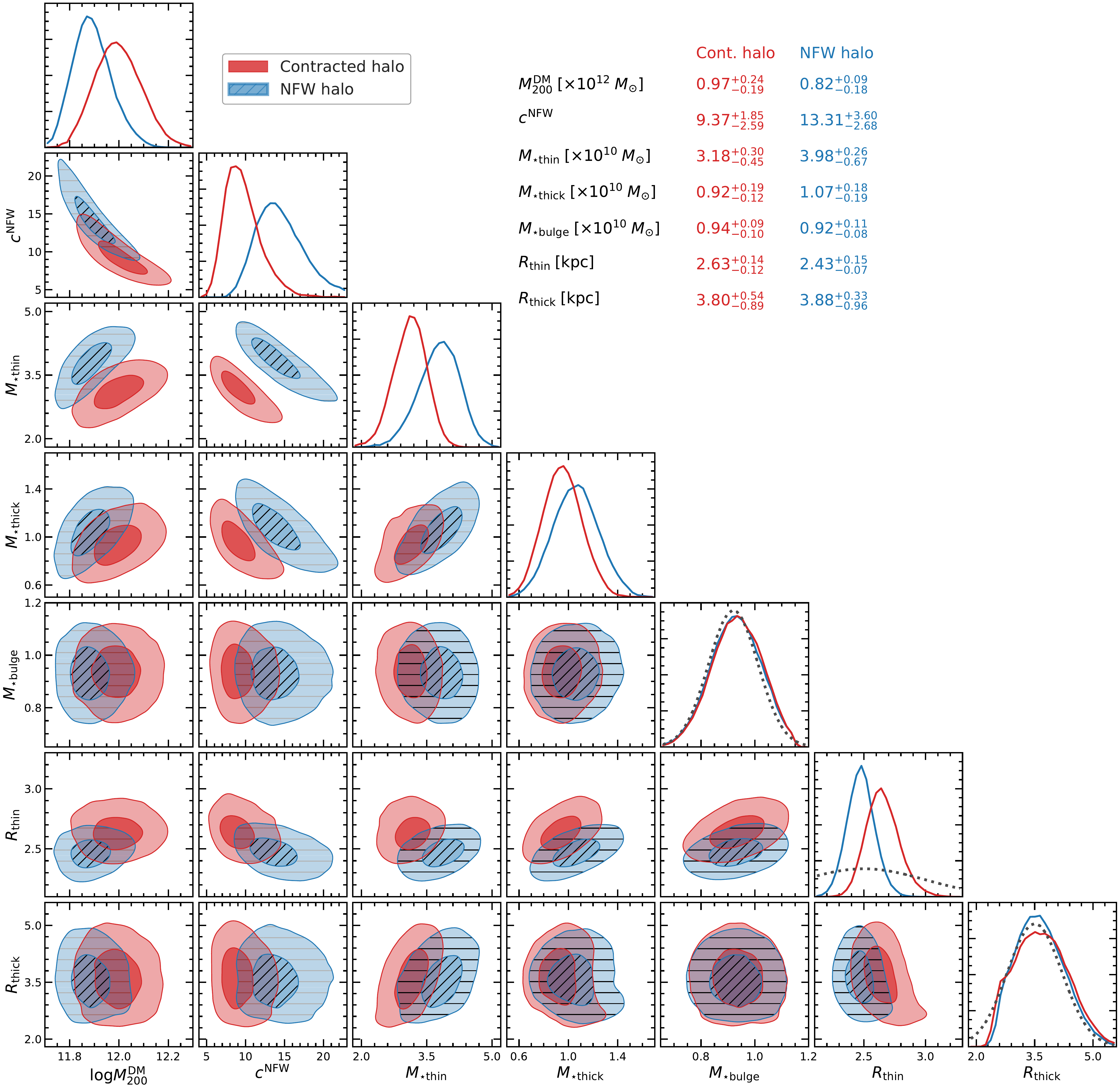}
        \vskip -.2cm
        \caption{ Posterior distributions pf the seven parameters of
          the MW mass model. The red shaded regions correspond to the
          contracted NFW halo model while the blue shaded regions
          correspond to a pure NFW halo. The darker and lighter contours
          enclose the 68 and 95 percentiles of the total probability
          respectively. The stellar masses are given in units of
          $10^{10}\Msun$ and the disc scale lengths in
          kiloparsecs. For convenience, the ML values as well as the
          68 percentile ranges are quoted in the top-right corner of
          the plot as well as in
          Table~\ref{tab:variable_parameters}. The grey dotted lines
          in the last three diagonal panels show the priors for the
          bulge mass, and the thin and thick disc scale lengths.  }
        \label{fig:MW_fit_corner_plot}
        \vskip -.2cm
\end{figure*}

To obtain the best fit model, we follow the Bayesian framework in
which the probability of a set of parameter values, $\theta=(\log
M_{200}^{\rm DM},c^{NFW},\rho_{0,\rm bulge},\Sigma_{0,\rm
  thin},\Sigma_{0,\rm thick},R_{\rm thin},R_{\rm thick})$, given the
data, $D$, is  
\begin{equation}
    p\left( \theta | D \right) = \frac{ p\left( D | \theta \right) \ p\left( \theta \right) }{ p\left( D \right) }
    \ ,
\end{equation}
where $p\left( D | \theta \right)$ is the probability of the data
given the model parameters, $p\left( \theta \right)$ is the prior
distribution of parameter values, and $p\left( D \right)$ is a
normalisation factor. We take three Gaussian priors for
$(\rho_{0,\rm bulge},R_{\rm thin},R_{\rm thick})$, as given in
the fourth column of Table~\ref{tab:variable_parameters}. For the
remaining parameters we consider no prior information; that is we take
a flat prior over a range much larger than the constraints inferred
from the data. The likelihood, $p\left( D | \theta \right)$, is taken
as the product of the likelihoods associated with each of the 41 data
points described in Section~\ref{subsec:MW_model:data}, that is 38
circular velocity measurements plus one data point for each of the following: the total
mass, thin to thick disc ratio, and the vertical force at the Sun's
position.
\changed{The circular velocity is calculated in the plane of the disc as $ V^2_{\rm circ} = R \ \rm{d} \Phi_{\rm tot}/\rm{d}R$, where $\Phi_{\rm tot}$ is the total gravitational potential and $R$ is the radial distance in the plane of the disc.}

We are interested in obtaining a global model that fits equally well
all the measurements within their uncertainties. However, when
considering only the observational errors for the \citet{Eilers2019}
rotation curve we find that the reduced $\chi^2$ is close to two and
that this large value is mostly driven by a couple of regions: a
dip in $V_{\rm circ}$ at $R\sim9\kpc$ and a second one at $R\sim20\kpc$ that are several $\sigma$ away
from the overall trend. Such outlying data points could drive the
model away from the set of parameters that give a good global fit and
force it to parameter values that better reproduce these local features,
even though such features are not expected to be captured by the
model. To mitigate any such problems, we 
\changed{introduce an additional model uncertainty, $\sigma_{\rm model}$, such that the total uncertainty associated with a data point is given by}
$\sigma=\sqrt{\sigma_{\rm obs}^2+ \sigma_{\rm model}^2}$, where $\sigma_{\rm obs}$ denotes the \citeauthor{Eilers2019} errors. 
\changed{We take $\sigma_{\rm model}=\mu\sigma_{\rm sys}$, where $\sigma_{\rm sys}$ is the systematic error associated with the \citeauthor{Eilers2019} determination. In Appendix \ref{appendix:good_fit} we compare different ways of defining $\sigma_{\rm model}$ and show that our results are largely insensitive to the definition of $\sigma_{\rm model}$.}
The quantity $\mu=0.21$ denotes a
weight factor whose value we have found by requiring that the reduced
$\chi^2$ should be unity (see Appendix \ref{appendix:good_fit} for details). Increasing the errors as discussed mostly
affects the points in the range $R\in[8,13]\kpc$ (the ones with very
small observational uncertainties of ${\sim}1\kms$) and leads to errors
that are at most a factor of 1.5 times higher than the observational
ones.

To find the best fitting model parameters and their associated
confidence intervals we employ a Markov Chain Monte Carlo (MCMC) approach
using the \textsc{emcee} python module \citep{Foreman-Mackey2013}. We
fit two different models for the DM halo: firstly, a profile that is 
contracted according to the baryon distribution, and, secondly, an
(uncontracted) pure NFW profile.

\subsection{The best fitting models}
\label{subsec:MW_model:best_fitting}

\subsubsection{The contracted halo model}
The best fitting MW rotation curve for the contracted NFW halo model is
shown as the solid red line in Figure~\ref{fig:MW_fit_data}. The black
data points are the \citet{Eilers2019} $\Vcirc$ data and the dark blue
square is the \citet{Callingham2019} total mass estimate converted to
a $\Vcirc$ value at the halo radius, $R_{200}$. The other colour data
points are the \citet{Posti2018}, \citet{Watkins2019} and
\citet{Eadie2019} estimates of the enclosed mass at various
Galactocentric radii, which were converted to circular velocities as 
$\sqrt{G M(<r)/r}$, where $G$ is Newton's gravitational constant
and $M(<r)$ is the enclosed mass within radius, $r$. The latter
measurements are inferred from the dynamics of globular clusters with
proper motions measured by \textit{Gaia} DR2 and several various HST
programs \citep[for details see][]{Eadie2019}.

Figure~\ref{fig:MW_fit_data} shows that the contracted NFW halo model
matches well the \citet{Eilers2019} and \citet{Callingham2019}
measurements, which were the ones used for the fitting procedure. The
model also agrees well with the mass measurements by \citet{Posti2018}
and \citet{Watkins2019}. However, it does not match the
\citet{Eadie2019} data particularly well, which may be due to the
assumption by these authors of a power-law model for the MW potential,
which is an oversimplification. For example, \citet{Eadie2018} have
tested their method against cosmological simulations and find that
their estimates are often affected by systematic uncertainties that
are not incorporated in their quoted error bars.
 
The good agreement between the model and the data can be clearly seen 
in the bottom panel of Figure~\ref{fig:MW_fit_data}, which shows the
difference between the predictions of the model and the various data
points. In particular, we notice two regions where the data show
systematic deviations from the model. Firstly, at $r\sim9\kpc$, the
data show a small, but statistically significant dip compared to the
model. This dip is probably due to localised irregularities in the
kinematics of our galaxy since it is also present when comparing
against a running average of the $\Vcirc$ data. Such local
irregularities are not allowed for in our global \Vcirc{} model and
thus it should not be surprising that the model does not reproduce
them. Secondly, at $r\sim20\kpc$, four neighbouring data points are
systematically 2-3$\sigma$ below the model predictions. This could be
a manifestation of systematic errors in the \citeauthor{Eilers2019}
\Vcirc{} data since the region $r\sim20\kpc$ is where some of their
model assumptions could break down (see their Figure~4). 

The best-fitting parameter values for the contracted NFW halo model are
given in the fifth column of Table~\ref{tab:variable_parameters} as
well as in the top right-hand corner of
Figure~\ref{fig:MW_fit_corner_plot}. The maximum likelihood (ML) model
corresponds to the MW residing in a DM halo with mass,
$M_{200}^{\rm DM}=0.97_{-0.19}^{+0.24}\times10^{12}\Msun$, and
concentration before baryon contraction,
$c^{\rm NFW}=9.4_{-2.6}^{+1.9}$. The ML value for the concentration
is, in fact, equal to the median concentration of
${\sim}1\times10^{12}\Msun$ haloes
\citep[e.g.][]{Neto2007,Hellwing2016}, implying that the MW resides in
an average concentration halo. Note that we did not use a prior for
the concentration and thus the very good agreement between our
inferred value and the theoretical predictions may be interpreted as a
validation that our model gives a good description of the Galactic
data.

The total mass of our galaxy is
$M_{200}^{\rm total}=1.08_{-0.14}^{+0.20} \times 10^{12}\Msun$, in
good agreement with the \citet{Callingham2019} measurement as well as
other mass determinations (see Figure 7 in
\citeauthor{Callingham2019}). As discussed previously, most of our
constraints come from the \Vcirc{} data and thus, even though we used
the \citeauthor{Callingham2019} value in our fit, the good agreement
of our $M_{200}^{\rm total}$ with this measurement is not
guaranteed.  Indeed, excluding the \citeauthor{Callingham2019}
measurement from our data sample does not introduce any systematic
differences in the inferred halo mass or concentration but results in
somewhat larger uncertainties.

We also find that the preferred MW stellar mass is
$M_{\star\ \rm total}=5.04_{-0.52}^{+0.43}\times10^{10}\Msun$, with
most (three fifths) residing in the thin disc and the remainder
equally split between the thick disc and the bulge (each containing
roughly one fifth of the total stellar mass). 
The constraints on the bulge mass are mostly given by the prior since the data we use, which corresponds to $R>5\kpc$, is largely insensitive to the mass or geometry of the bulge (see Figure~\ref{fig:MW_fit_corner_plot}).
Most of the baryonic
mass within the halo is in the gaseous component:
$1.2\times10^{10}\Msun$ as \HI{} and molecular gas, and
$6.4\times10^{10}\Msun$ as the CGM. Adding up everything, we find that
the MW contains roughly 72\% of the cosmic baryonic fraction. Caution
should be taken when interpreting this result since the cold gas and
especially the CGM distribution in the MW are rather uncertain. Here,
we have modelled the CGM using the average predictions from
hydrodynamical simulations, not taking into account halo-to-halo
variation in CGM mass, which the simulations predict is rather large.

The contribution of the various MW components to the total rotation
curve of the best-fitting model is shown in
Figure~\ref{fig:MW_best_fit_components}. The shaded regions around
each curve show the 68 percentile confidence intervals. The inner
region, $R<10\kpc$, is dominated by baryons, in particular by the
stellar component. Our inferred stellar mass is slightly smaller than
the \citet{Bovy2013a} estimate, but consistent within the $68$
percentile errors (see black symbols with error bars). However, we
find a much more massive DM halo than \citeauthor{Bovy2013a}. This is
mostly the result of the latest \textit{Gaia} data which favour a MW
rotation curve of $(229\pm1)\kms$ at the Solar position, rather than
the $(218\pm10)\kms$ value inferred by \citeauthor{Bovy2013a}.  Our
results also solve a long-standing puzzle: previous measurements
suggested that the MW rotation curve is dominated by the
stellar component up to distances of $R{\sim}12-14\kpc$
\citep[e.g.][]{Bovy2013a,Eilers2019}, in disagreement with recent
hydrodynamical simulations that find that the DM should already be
dominant for $R>5\kpc$
\citep[e.g.][]{Schaller2015a,Grand2017,Lovell2018a}. In our model, the
Galactic DM halo exceeds the stellar component contribution at
$R{\sim}8\kpc$, in good agreement with the theoretical predictions
(see Figure~11 in \citealt{Lovell2018a}) when accounting for the fact
that the MW is a 1$\sigma$ outlier in the stellar-to-halo mass
relation (see discussion in
Section~\ref{subsec:discussion:which_is_better}).

To test the effect of the CGM, we have considered two variants of our
MW model: i) excluding a CGM component altogether, and ii) assuming
that the CGM mass is nearly twice as large as in the fiducial model
such that the MW halo contains the universal baryonic fraction. In
both cases the CGM contribution to the rotation curve is negligible
for $r\lesssim30\kpc$ and hardly affects the best-fit values of the
stellar discs or the DM halo. The largest effect is on the total mass
of the MW and even then the variation is small, well within the quoted
uncertainty range (the total mass increases by $5\%$ in the model
with the most massive CGM component compared to the model without a
CGM).

To get a better understanding of the various degeneracies
between the model parameters, we show in
Figure~\ref{fig:MW_fit_corner_plot} the posterior distribution for
each pair of parameters. In the off-diagonal panels, the red shaded
regions illustrate the 68 and 95\% confidence regions, while, in the
diagonal panels, the red lines show the marginalised probability of
each model parameter. To aid the physical interpretation, we have
converted the bulge and the stellar disc densities, which are the
parameters used in the fitting procedure, to the total stellar mass of
the bulge, thin and thick disc, and only show these quantities in
Figure~\ref{fig:MW_fit_corner_plot}.

Figure~\ref{fig:MW_fit_corner_plot} shows that most parameters are
weakly correlated but there are a few interesting degeneracies. Most
pronounced is the degeneracy between DM halo mass and
concentration. As we already discussed, most of the model constraints
come from the inner regions, i.e. $r\lesssim20\kpc$, and the same
enclosed mass can be obtained by, for example, decreasing the halo
concentration in tandem with increasing the halo mass. We also find a
positive correlation between halo mass and the thin and thick disc
stellar masses: more massive haloes prefer a more massive stellar
disc. This is because a more massive halo provides a similarly good
fit only if it has a lower concentration, and thus has less mass in
the inner region, which, in turn, can be compensated for by a larger
disc mass. The same effect explains the negative correlation between
halo concentration and disc masses, and the positive correlation
between the thin and thick disc masses.

\subsubsection{The pure NFW halo model}
As we discussed extensively in Section~\ref{sec:halo_contraction}, the
accretion and settling of baryons causes a contraction of the DM halo
density. Many previous studies have neglected this contraction and
instead have assumed that the halo is still well fit by an NFW
profile. To understand any systematic effects arising from this
incorrect assumption, we proceed to fit also an NFW halo model to the
same data sample.

The best fitting NFW halo model is shown by the dashed blue line in
Figure~\ref{fig:MW_fit_data}. We find that this model fits the data
almost as well as the contracted NFW halo model (we discuss this in detail
in the next subsection). In particular, in the range $r \in
[4,50]\kpc$ the difference between the \Vcirc{} predictions of the two
models is less than $1\kms$. However, the best fitting NFW model has
very different parameters values than the contracted halo model. The
best fitting NFW halo has a lower mass, $M_{200}^{\rm
  DM}=0.82_{-0.18}^{+0.09}\times10^{12}\Msun$, and a higher
concentration, $c^{\rm NFW}=13.3_{-2.7}^{+3.6}$.  

As we have shown in Figure~\ref{fig:MW_density profile}, the
contracted NFW halo corresponding to the observed baryonic mass
distribution of the MW is not well described by an NFW profile, which
raises the question: How can the NFW halo model give as good a fit to
the \Vcirc{} data as the contracted halo model? The answer lies in the parameters describing the
baryonic component of the MW, which have different values in the two models. For the pure NFW halo model, the MW total
stellar mass is $5.97_{-0.80}^{+0.40}\times10^{10}\Msun$, roughly 20\%
higher than in the contracted halo model, and, furthermore, the baryon
distribution is somewhat more concentrated, with the thin disc scale
length smaller in the NFW halo case. All these differences can be
gauged from Figure~\ref{fig:MW_fit_corner_plot}, which contrast the
inferred parameters in the two models. 

\vspace{-.2cm}
\section{Discussion}
\label{sec:discussion}
In this work we have introduced a phenomenological approach to
describe the density profile of a halo that has been modified by
baryons settling at its centre (see
Section~\ref{sec:halo_contraction}). When applied to our Galaxy, the
halo contraction model predicts that the inner region contains far
more DM than would have been the case in the absence of baryons. The
inner regions, $r{\sim}1\kpc$, see a substantial increase in
enclosed DM mass while at the Sun's position the factor is
${\sim}2$. The exact numbers depend on both the concentration and the
mass of the DM halo in which our galaxy has formed: haloes with lower
concentration or lower mass experience a larger contraction.

That baryons can cause a DM halo to contract has been known for a long
time \citep{Blumenthal1986,Barnes1987}, and this has been confirmed by
many subsequent studies
\citep[e.g.][]{Gnedin2004,Abadi2010,Duffy2010,Schaller2015a,Dutton2016}. However,
Galactic studies often neglect this effect and model the DM halo as an
NFW profile. While the NFW formula gives an excellent description of
the DM radial density profile in DM-only simulations, it cannot
capture the changes induced by the baryonic distribution. As we have
shown in Section~\ref{subsec:MW_bias_halo_properties}, incorrectly
describing the DM halo as an NFW profile leads to biases in the
inferred halo mass and concentration. These biases are largest when
modelling the enclosed mass at Galactocentric distances $\leq10\kpc$;
however they are non-negligible even at larger distances, or when
modelling escape velocity measurements.

Using the latest Galactic rotation curve data together with a few
other measurements, we have fitted a Galactic model with seven
parameters, two characterizing a spherically symmetric DM halo and five
the MW stellar distribution. We have found that the MW rotation curve
is very well described by a contracted NFW halo with a mass of
$1\times10^{12}\Msun$ and an original (i.e. before baryonic
contraction) concentration of $9$, which is in remarkable agreement
with the halo mass--concentration relation predicted by $\Lambda$CDM
cosmological simulations \citep{Navarro1997,Neto2007}. Furthermore, our results are in very good agreement with the \citet{Li2019a} recent determinations of the DM halo mass and concentration, which are based on the dynamics of MW satellites that are far enough from the Galaxy such that baryonic effects can be neglected. 

The same data are also well described by a pure NFW halo but of mass,
$0.8\times10^{12}\Msun$ and concentration of $13$.
To fit the data, the pure NFW halo model requires the MW to have a more
massive stellar disc, ${\sim}5\times10^{10}\Msun$, than inferred from
the contracted halo model, ${\sim}4\times10^{10}\Msun$. This 25\%
discrepancy is illustrated in the top panel of
Figure~\ref{fig:discussion_observables_comparison}. Currently, the
stellar disc mass of our galaxy is poorly constrained \citep[e.g. see
the compilation of stellar profiles in][]{Iocco2015} and thus cannot
be used to distinguish between the two models, although most
measurements agree better with the lower stellar mass of the
contracted NFW halo model
\citep[e.g][]{Bovy2013a,Bland-Hawthorn2016}. This raises an intriguing
question of what kind of existing or forthcoming data can distinguish
between the contracted and the pure NFW models of the Galactic DM
halo? We now address this question.

\subsection{Which Galactic halo model is better: a contracted NFW or a
  pure NFW profile?}
\label{subsec:discussion:which_is_better}

\begin{figure}
        \centering
        \includegraphics[width=.9\linewidth,angle=0]{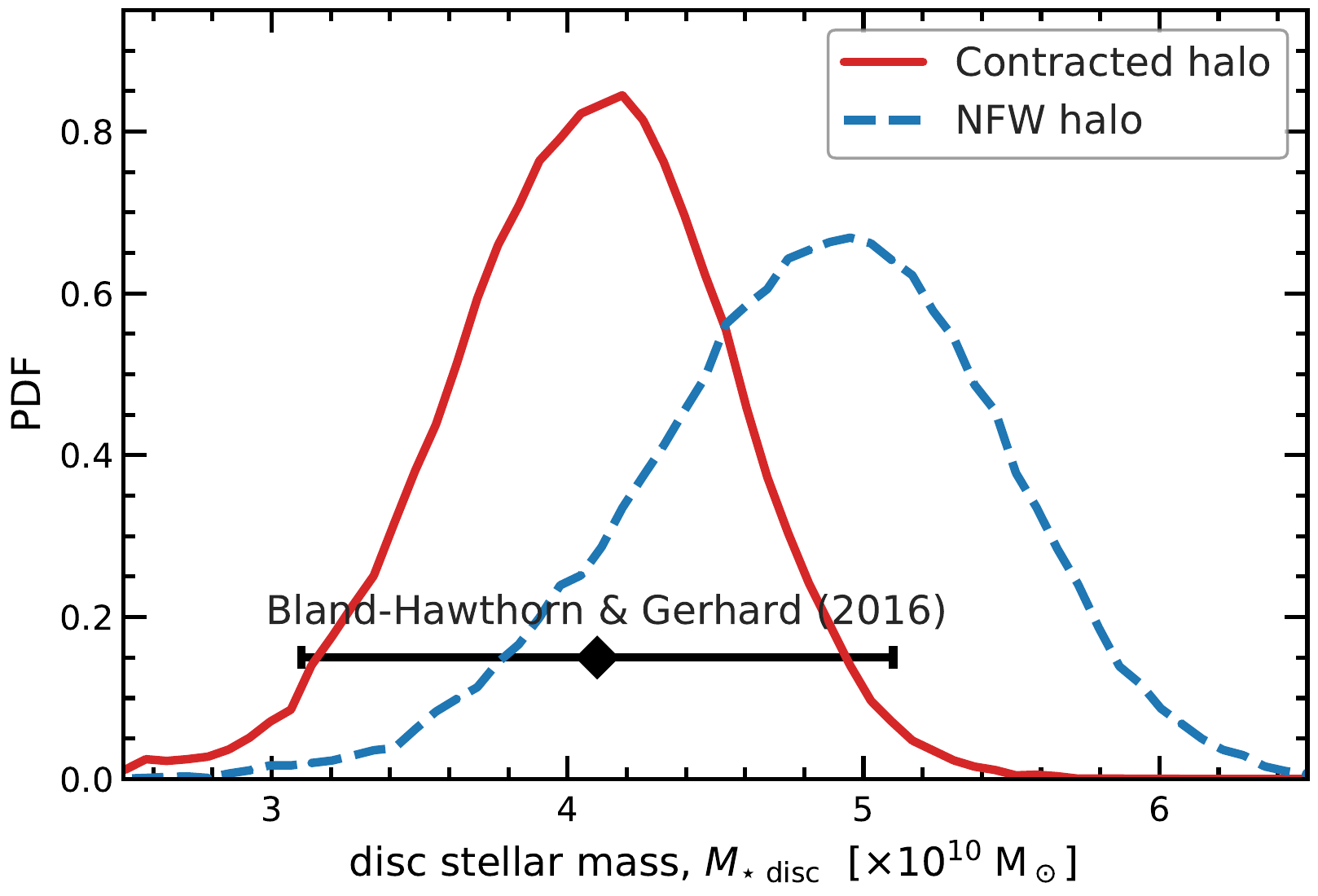}
        \includegraphics[width=.9\linewidth,angle=0]{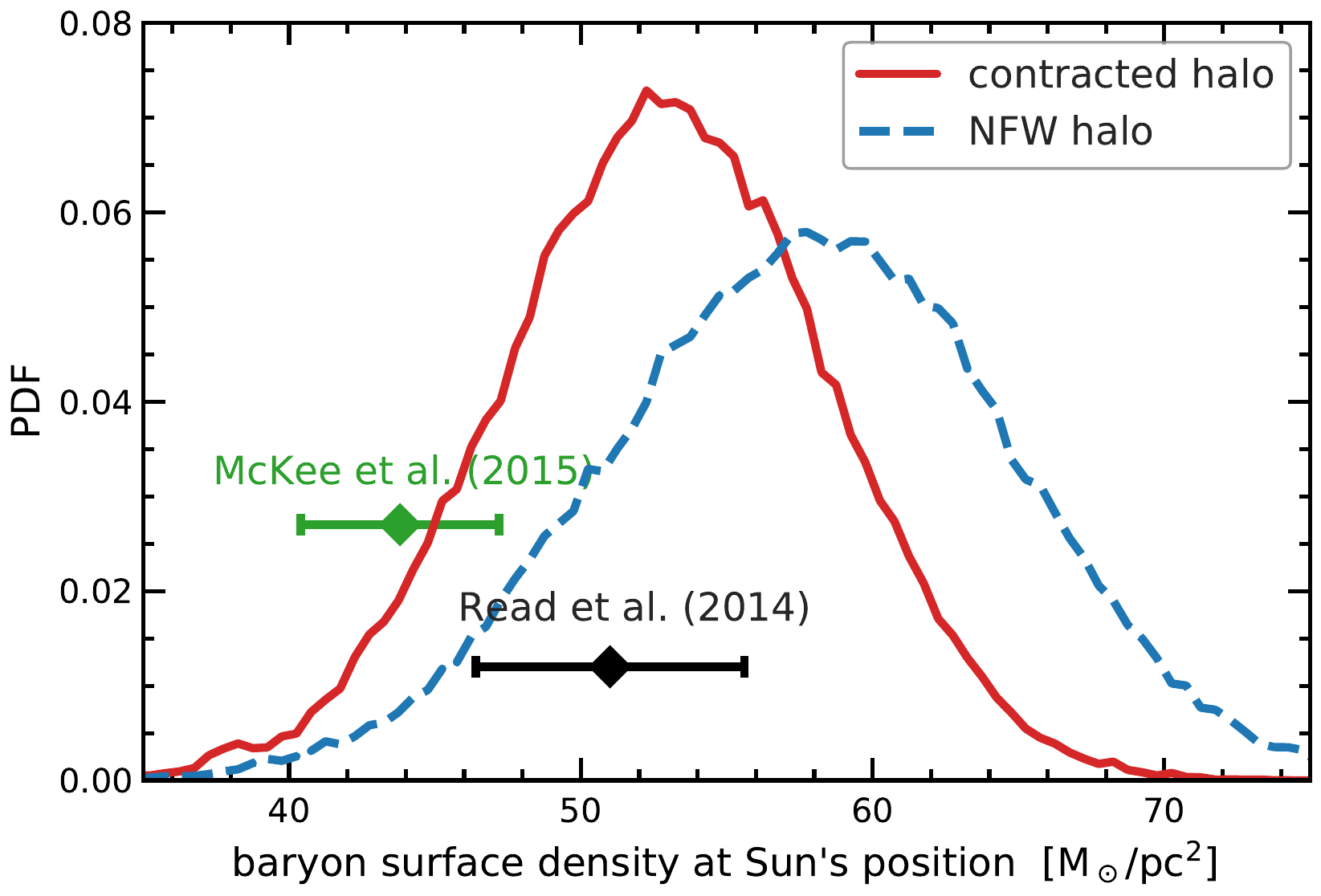}
        \includegraphics[width=.9\linewidth,angle=0]{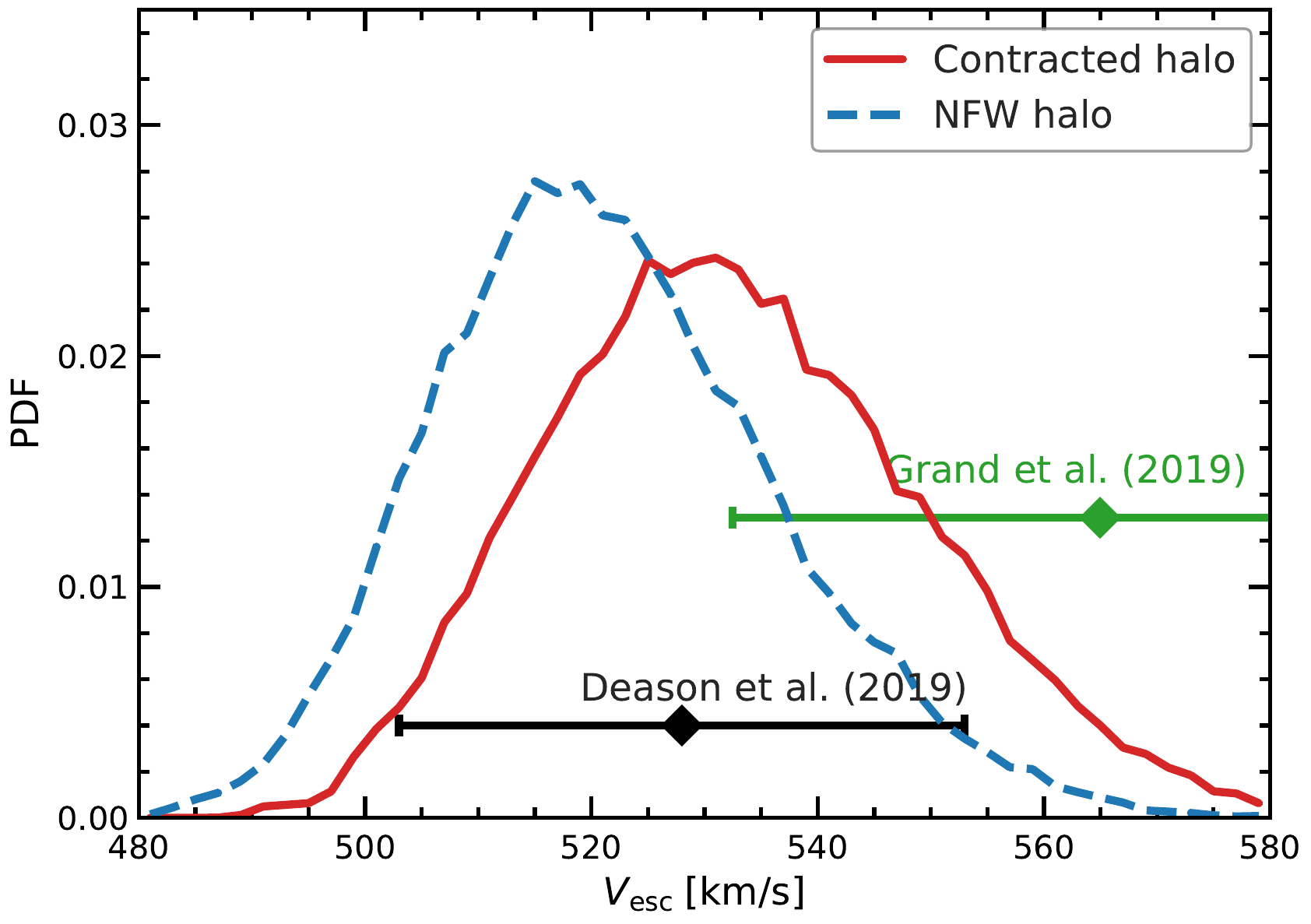}
        \vskip -.2cm
        \caption{ Examples of observables that may be used to
          constrain the description of the Galactic DM halo as either
          a contracted NFW or a pure NFW profile. The solid red and
          blue dashed lines show the marginalised probabilities of the
          observables inferred from fitting the MW rotation curve with
          contracted and pure NFW profiles, respectively. The three
          panels show: disc stellar mass (top), baryonic surface
          density within $1.1\kpc$ from the disc at the Solar position
          (middle), and escape velocity at the Solar position
          (bottom). The diamonds with horizontal error bars show
          recent measurements and their 68\% confidence limits:
          \citet{Bland-Hawthorn2016} estimate of disc stellar mass
          based on a compilation of studies, \citet{Read2014} and
          \citet{McKee2015} estimates of the baryonic density at the
          Sun's position, and the recent escape velocity measurement
          of \citet{Deason2019} and the updated value of
          \citet{Grand2019a}.  }
        \vskip -.2cm
        \label{fig:discussion_observables_comparison}
\end{figure}

Within the standard $\Lambda$CDM paradigm we expect, both from
theoretical considerations and cosmological simulations, that the
contracted halo is the physically motivated model. Nonetheless, it is
desirable to check the extent to which this model is favoured by the
observations.

Visually, we find that both the NFW and the contracted NFW halo models
give a good fit to the MW observations analyzed in this study (see
Figure~\ref{fig:MW_fit_data}). The goodness of fit can be better
quantified by the likelihood of the best fitting model, and,
in particular, by comparing the maximum likelihood of the two
models. Both have the same number of free parameters and thus
comparing them is straightforward. We find that the contracted NFW
halo model is slightly preferred since it has a maximum likelihood
that is a factor of 2.1 times larger than the pure NFW halo model,
corresponding to a p-value of 0.48. Thus, while the contracted NFW
halo model is better at describing the data, the difference is too
small to rule out the pure NFW halo model.

Figure~\ref{fig:MW_fit_data} shows that while the best fitting
contracted and pure NFW halo models have the same rotation curve in
the region $r\in[5,60]\kpc$, they predict different behaviours outside
this region. For example, for $r<2\kpc$, the pure NFW model predicts a
rotation curve that is systematically lower by $10\kms$ and thus
potentially this region can be used to distinguish the two
models. However, current \Vcirc{} data do not constrain the bulge
mass, which in our model is mostly determined by the prior, and thus
it is conceivable that, by preferring slightly different stellar bulge
masses, both models could predict equal \Vcirc{} values at $r<5\kpc$.

The contracted and pure NFW halo models also predict different
\Vcirc{} values at large Galactocentric distances. At 200\kpc{} the
contracted halo predicts a 5\kms{} (${\sim}3\%$) higher rotation
velocity than the pure NFW, which potentially can be used to
distinguish between the two. Current measurements at that distance are
not yet accurate enough, since, for example, the
\citealt{Callingham2019} mass measurement has a 15\% uncertainty which
translates into a $7.5\%$ error in \Vcirc{}. The mass uncertainties
could be reduced to the $10\%$ level ($5\%$ in \Vcirc{}) when accurate
proper motions become available for most of the ultra-faint MW
satellites (see Figure~11 in \citealt{Callingham2019}) and could be
reduced even further by combining with other halo tracers such as
globular clusters and halo stars.

The MW CGM is still uncertain and assuming different CGM masses could
decrease the discrepancy between the models. For example, if the MW
halo contained the universal baryon fraction, within $200\kpc$ we
would expect a baryonic mass of $12.5\times10^{10}\Msun$, of which
slightly more than half is in the form of stars and cold dense gas at
the centre of our galaxy (see
Table~\ref{tab:variable_parameters}). Thus, by varying the CGM mass
from zero to its maximum allowed value (the universal baryon fraction;
it is unlikely that a halo could contain many more baryons than the
mean cosmic fraction), \Vcirc{} can vary by up to 4\% at
$r=200\kpc$. This variation is equal to the predicted difference
between the contracted and pure NFW models at that distance and it is an important systematic that needs to be accounted for.

The best fitting contracted and pure NFW halo models imply different
masses for the Galactic stellar disc, and one way to test for this is
by comparing the baryonic surface density at the Solar position. In
the middle panel of Figure~\ref{fig:discussion_observables_comparison}
we show the total baryon projected density within $1.1\kpc$ from the
disc plane. The contracted NFW halo model predicts a surface density
that is systematically lower (by nearly 20\%) than the NFW halo
model. The two recent determinations of \citet{Read2014} and
\citet{McKee2015} favour the contracted halo model; however, due
to large uncertainties, the pure NFW model cannot be ruled out.

The escape velocity at the Solar location can also be used to
differentiate between the two models, as illustrated in the bottom
panel of Figure~\ref{fig:discussion_observables_comparison}. Although
the two $V_{\rm esc}$ distributions overlap, the contracted NFW halo
model predicts a $V_{\rm esc}$ value that is systematically higher by
${\sim}10\kms$. Current $V_{\rm esc}$ measurements are not precise
enough to differentiate between the two models, although the
\citet{Grand2019a} value, which is an update of the \citet{Deason2019}
measurement accounting for systematics such as halo substructure and
stellar halo assembly history, favours the contracted halo model.

\begin{table}
    \centering
    \caption{ Summary of observables and measurements that can be used
      to choose between a contracted and a pure NFW profile as the
      best description of the Galactic DM halo. None of the
      measurements can yet be used to rule out either of the models,
      so here we show which of the two is preferred by each
      measurement, which is indicated by the \checkmark{} symbol. The
      last column of gives the ratio of likelihoods between the
      contracted and pure NFW halos for each measurement (a value
      larger than unity means that the contracted NFW halo model is
      preferred).  }
   
    \renewcommand{\arraystretch}{1.1} 
    \begin{tabular}{ p{.4\linewidth} p{.1\linewidth} p{.07\linewidth} p{.07\linewidth} p{.1\linewidth} } 
        \hline\hline
        Observable & Study &  Cont. halo & NFW halo & $\mathcal{L}$ ratio \\
        \hline  \\[-.2cm]
        
        Theoretical predictions$^\dagger$ & -- & \checkmark & & -- \\ 
        \rowcolor{Gray}
        Fit to MW rotation curve & (1) & \checkmark & & 2.1 \\ 
        Stellar disc mass & (2) & \checkmark & & 1.3 \\
        \rowcolor{Gray}
        Abundance matching & (3) & \checkmark & & 2.8 \\
        \rowcolor{Gray}
          & (4) & \checkmark & & 1.9 \\
        Baryon surface density & (5) & \checkmark & & 1.6 \\
        at Solar position  & (6) & \checkmark & & 2.3 \\
        \rowcolor{Gray}
        Escape velocity & (7) & \checkmark & & 1.1 \\
        \rowcolor{Gray}
        at Solar position  & (8) & \checkmark & & 1.3 \\
        \hline\hline
    \end{tabular}
    
    \vskip .1cm
    \renewcommand{\arraystretch}{1.2}
     \begin{tabular}{ @{} p{1\columnwidth} @{} }
        References: (1) this work, (2) \citet{Bland-Hawthorn2016}, (3) \citet{Moster2013}, (4) \citet{Behroozi2013}, (5) \citet{Read2014}, (6) \citet{McKee2015}, (7) \citet{Deason2019}, (8) \citet{Grand2019a}. \\
        $^\dagger$ Many hydrodynamical simulations find that the DM
       halo profile changes in the presence of baryons
       \citep[e.g.][]{Gnedin2004,Abadi2010,Duffy2010,Schaller2015a,Dutton2016}. \\[-.2cm]
    \end{tabular}
    \renewcommand{\arraystretch}{1.0}
    \label{tab:discussion_which_model_is_preffered}
\end{table}

Another way to differentiate between the two halo models is to compare
them with the stellar to halo mass relation. This is a specially
powerful test since the pure NFW halo model predicts a lower total mass 
but a higher stellar mass than the contracted NFW halo model. Using
the \citet{Moster2013} abundance matching results, we find that, for
the contracted NFW halo model, the MW stellar mass is 0.13 dex above the
mean trend ($0.9\sigma$ away). In contrast, for the pure NFW halo the
stellar mass is 0.26 dex higher than the mean, a $1.7\sigma$
outlier. We obtain a similar result if instead we consider the
\citet{Behroozi2013} abundance matching relation, with the MW stellar
mass being $0.8$ and $1.4\sigma$ above the median trend for the
contracted and pure NFW halo models, respectively. The main difference
between the \citeauthor{Moster2013} and \citeauthor{Behroozi2013}
relations is that the latter has a larger scatter in the stellar mass
at fixed halo mass (0.15 dex versus 0.22 dex).  Thus, comparison with
the stellar to halo mass relation also favours the contracted halo
model but is not conclusive. 

In Table~\ref{tab:discussion_which_model_is_preffered} we provide a
summary of the observables we just discussed and study the extent to
which various Galactic measurements favour either the contracted or
the pure NFW halo models. We calculate the joint likelihood of the
measured values (assuming Gaussian uncertainties) and compare with our
predictions for those observables inferred using the contracted and
pure NFW halo models. In all cases, we find that the contracted halo
model is preferred, but due to the large uncertainties, the
differences are rather modest. One way to discriminate between the two
models is to calculate the joint probability of the measurements shown
in Table~\ref{tab:discussion_which_model_is_preffered}. To be
conservative, for each observable that has more than one entry in the
table, e.g. abundance matching, we choose the entry that discriminates
the least between the models. We find that the contracted NFW halo
model has a 9 times higher likelihood (p-value of 0.11) than the pure
NFW one.

\subsection{DM density at the Solar position}
\label{subsec:discussion:solar_DM}

\begin{figure}
        \centering
        \includegraphics[width=.9\linewidth,angle=0]{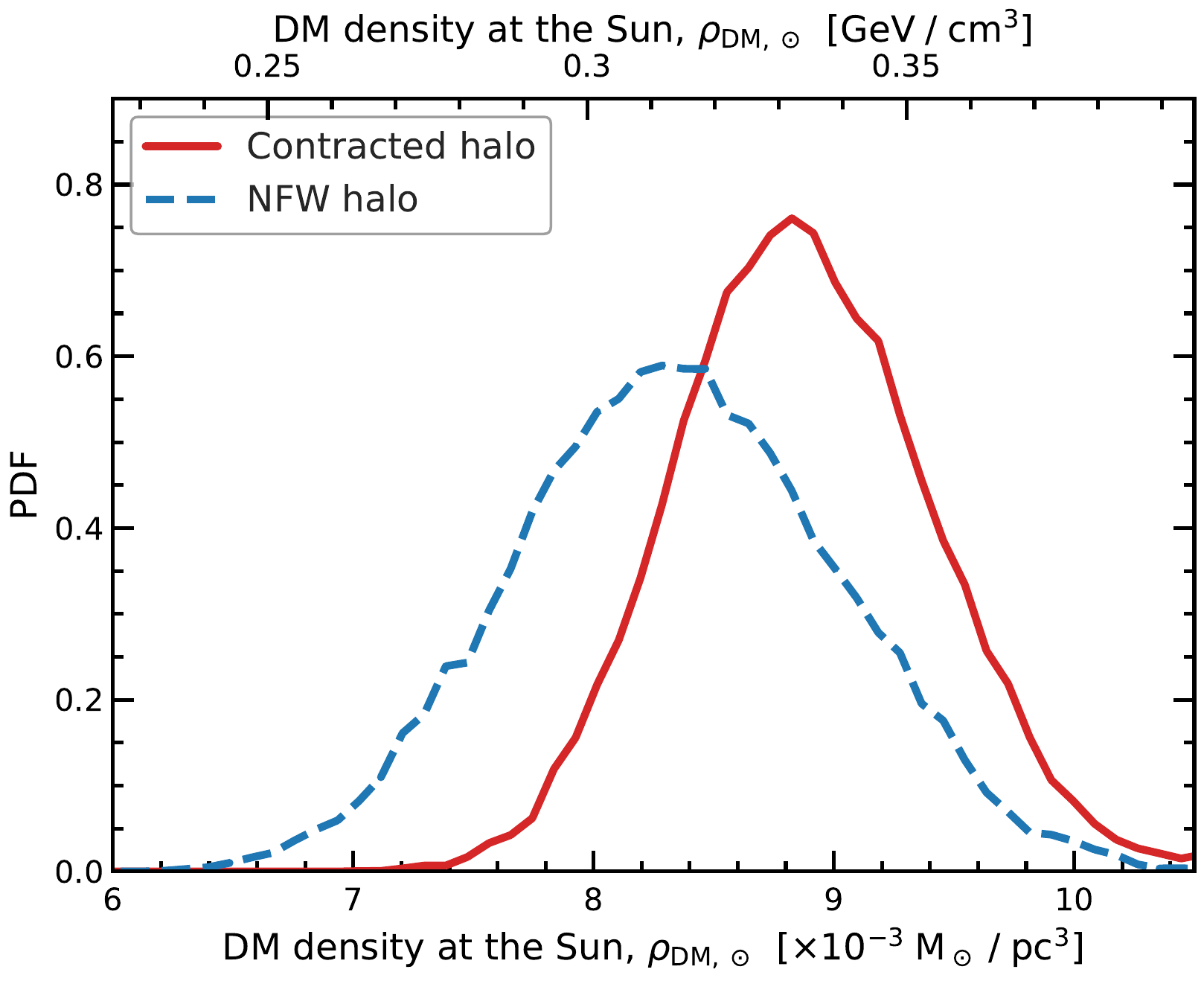}
        \vskip -.2cm
        \caption{ The DM density at the position of the Sun derived
          when modelling the Galactic DM halo as a contracted (solid
          red) or as a pure NFW halo (blue dashed). Modelling the MW
          as the contracted halo results in a $10\%$ higher DM density
          that in the pure NFW halo case.  }
          \label{fig:discussion_DM_density_Sun}
\end{figure}

One of the key products of Galactic mass models is the local density
of DM, which is important for direct detection experiments. The
inferred local DM density given by our model is shown in
Figure~\ref{fig:discussion_DM_density_Sun}, where the solid and dashed
lines correspond to the contracted and  pure NFW halo models,
respectively. The contracted halo model indicates a local DM density
of $8.8_{-0.5}^{+0.5}\times10^{-3}\Msun\pc^{-3}$, that is,
$0.33_{-0.02}^{+0.02}~\rm{GeV}~\rm{cm}^{-3}$, in agreement with other
literature values \citep[e.g. see Figure 1 in the review
by][]{Read2014}. The NFW halo model predicts a DM density that is
systematically lower than this by 10\%, which is due to the fact that
the baryonic disc is more massive in that case and thus accounts for a
larger fraction of the matter distribution at the Solar position. This
result supports previous studies that have found that the poorly known
baryonic distribution in the MW is the main systematic uncertainty in
the determination of the local DM density
\citep[][]{Buch2019,Karukes2019,deSalas2019}.
\changed{For direct detection experiments, the poorly constrained effect of baryons on the tail of the DM velocity distribution can lead to an even higher uncertainty \citep[][]{Bozorgnia2016,Bozorgnia2019,Kelso2016,Sloane2016,Callingham2019b}.}

The comparison between the contracted and pure NFW halo models
highlights the desirability of using a physically motivated global
model for our Galaxy. Often, for example as we have found in
Figure~\ref{fig:MW_fit_data}, the data can be equally well fitted by
several models that are degenerate in the properties of the baryonic
and the DM components. In such cases, hydrodynamical simulations
provide an important guide by offering plausible arguments why certain
models are to be preferred and thus can help break the degeneracy
between the baryon and DM distributions. Our study illustrates the
systematic biases in the inferred local DM distribution that can be
introduced by incorrectly modelling the halo using a pure NFW
profile. Biases are also likely to be present when modelling the MW
halo as a gNFW profile, since this functional form is not flexible
enough to capture the contracted DM halo profile (see bottom panel of
Figure~\ref{fig:MW_density profile}).

\vspace{-.3cm}
\subsection{The total mass of our galaxy}
\label{subsec:discussion:total_mass}
For the contracted halo model we find that the total mass of the MW
within a radius enclosing a mean density of 200 times the critical
density is
$M_{200}^{\rm total} = 1.08_{-0.14}^{+0.20}\times10^{12}\Msun$, in
good agreement with many recent measurements based on the
\textit{Gaia} DR2 data
\citep[e.g.][]{Posti2018,Eadie2019,Li2019a,Watkins2019,Karukes2020,Wang2020}. While our method uses
the \citet{Callingham2019} total mass measurement as one of the data
points to which we fit our model, we infer roughly the same total
mass if we remove the \citeauthor{Callingham2019} data point (although
with somewhat larger uncertainties). Our determination may thus  be
seen as a largely independent constraint on the MW total mass.

Our results also highlight that the total mass estimate is sensitive
to systematic uncertainties arising from the modelling of the DM
halo. Depending on which measurement is being considered, incorrectly
modelling the MW halo as an uncontracted NFW profile can both
overestimate or underestimate the total mass. For example, modelling
the enclosed mass within a fixed Galactocentric distance as a pure NFW
profile with the typical halo mass--concentration relation leads to an
overestimate of the total mass (see
Figure~\ref{fig:constraint_enclosed_mass}; the same holds true of the
escape velocity measurement but the systematic error in this case is
lower -- see Figure~\ref{fig:constraint_escape_velocity}). In
contrast, modelling the entire rotation curve as an NFW profile leads
to an underestimation of the total mass (see
Table~\ref{tab:variable_parameters} and
Figure~\ref{fig:MW_fit_corner_plot}). This is because to account for
the baryon-induced DM halo contraction, the data prefer a high
concentration for the NFW profile which, given the halo
mass--concentration degeneracy in the modelling, results in too low a
DM halo mass. This potentially explains why mass estimates based on
fitting the rotation curve \citep[e.g.][]{Bovy2012a,Kafle2014} are
systematically lower than determinations based on other methods
\citep[e.g. see the comparison in][]{Wang2015a}.

\vspace{-.3cm}
\subsection{Limitations and future improvements}
\label{subsec:discussion:limitations}
Our model assumes a spherically symmetric DM halo but cosmological
simulations predict ellipsoidal shapes
\citep[e.g.][]{Frenk1988,Bett2007,Schneider2012}. This simplification
is unlikely to affect our results since the baryonic distribution
leads to a roughly spherical DM distribution in the inner regions,
i.e. for $r\lesssim20\kpc$, which is the region where the best quality
rotation curve data exists \citep{Gnedin2004,Abadi2010}. An
approximately spherical shape for the inner Galactic halo is also
supported by observational data \citep[e.g.][]{Posti2018,Wegg2019}. At
larger distances, the flattening of the DM halo becomes important and
can affect the dynamics of halo tracers
\citep[e.g.][]{Law2010,Bowden2013,Shao2019a}. However, we have used
only one measurement at such distances, the \citet{Callingham2019}
total mass estimate, which is inferred under the assumption of
spherical symmetry, with deviations from this assumption being
accounted for in the uncertainties and, thus, unlikely to bias our
model estimates.

In fact, having a non-spherical DM halo introduces an entire new layer
of complexity since both the flattening and the orientation of the MW
DM halo can vary with radial distance
\citep[e.g.][]{Bailin_2005,Shao2016}. In particular, based on
hydrodynamical simulations the inner halo is expected to be aligned
with the Galactic disc, while the orbit of the Sagittarius stream, as
well as the disc of satellite galaxies, indicate that the outer halo
is perpendicular to the MW disc
\citep{Law2010,Vera-Ciro2013,Shao2019b}, with the transition between
the two halo orientations occurring at an as yet unconstrained distance.

As the MW data become ever more abundant and accurate, deviations from
the smooth (i.e. featureless) stellar disc and halo model used here
can become increasingly important. Such deviations can arise from the
dynamics of the spiral arms \citep[e.g.][]{Kawata2018,Hunt2018}, perturbations to the disc from the
Sagittarius and LMC dwarfs \citep[e.g.][]{Gomez2017a,Laporte2018}, or
from departures of the DM halo from equilibrium due to the recent
accretion of the LMC
\citep[e.g.][]{Erkal2018,Cautun2019a,Garavito-Camargo2019}, which is
thought to be significantly massive, with a total mass at infall of
${\sim}2.5\times10^{11}\Msun$ \citep{Penarrubia2016,Shao2018b}. In
fact, there is a dip in the MW rotation curve at ${\sim}9\kpc$ from
the Galactic Centre that is a several sigma outlier from the mean
predictions of both the contracted and pure NFW halo models. To
accommodate it, we 
\changed{introduced an additional model uncertainty when fitting the}
rotation curve data. However, this approach potentially downgrades the
constraining power of the data and a better way forward would be to
identify the physical cause of the deviation and model it.

\section{Conclusions}
\label{sec:conclusion}

We have determined the Galactic DM and baryon mass profiles using the
latest \textit{Gaia} DR2 rotation curve. We modelled the baryon
distribution assuming six components: a bulge, thin and thick stellar
discs, \HI{} and molecular gas discs and a CGM. The DM halo was
modelled as an NFW profile that has been contracted by the
accumulation of baryons, using a prescription calibrated on the latest
hydrodynamical simulations of MW-mass haloes. Throughout the paper we
contrasted the results of this contracted halo model with the common
approach taken in the literature of neglecting the baryon-induced
contraction of the DM halo.

We first investigated the effect that baryons have on the DM
distribution using three recent sets of hydrodynamical simulations of
MW-mass halos: \auriga{}, \apostle{} and \eagle{}. All of them show
that the addition of baryons modifies the DM halo profile predicted by
DM-only simulations and that the effect is largest at distances
$r<10\kpc{}$, where the enclosed DM mass can be a factor of a few to
several times higher than in the absence of baryons. The change in the
DM halo profile can be expressed in terms of a non-linear relation
between the DM and total mass ratios (see
Eq. \ref{eq:DM_increase_fit}) that is consistent across our three
simulation sets (see Figure~\ref{fig:DM_fraction} and
\ref{fig:method_check}), and that roughly agrees with the
\citet{Blumenthal1986} and \citet{Gnedin2004} adiabatic contraction
approximations, although we do find systematic deviations at the 10\%
level.

We studied the baryon-induced contraction of the Galactic DM halo  to
find that:
\begin{itemize}
\item Compared to the expectation from DM-only simulations, the
  baryons increase the enclosed DM mass by a factor of roughly $1.3$,
  $2$ and $4$ times at radial distances of $20$, $8$ and
  $1\kpc$ respectively (see Figure~\ref{fig:MW_mass_fraction}).
    \item For a fixed baryonic mass, the amplitude of the contraction
      depends on the mass and concentration of the original
      (uncontracted) halo, and is larger for lower mass and lower
      concentration haloes (see Figure~\ref{fig:MW_mass_fraction} and
      \ref{fig:MW_enclosed_mass}). 
    \item The contracted DM density profile of the MW varies as
      $r^{-2}$ over a wide range of distance, $r\in[5,30]\kpc$.  The
      contracted profile cannot be described by NFW, gNFW or Einasto
      profiles (see Figure~\ref{fig:MW_density profile}).
    \item Incorrectly modelling the MW halo as a pure NFW profile
      results in systematic biases in the inferred mass and
      concentration of the halo (see
      Figure~\ref{fig:constraint_enclosed_mass} and
      \ref{fig:constraint_escape_velocity}). These biases are present
      for both enclosed mass and escape velocity measurements and are
      largest at small $r$ where the halo contraction is largest.
\end{itemize}

\vskip .2cm

Finally, we fitted the MW mass model to the \textit{Gaia} DR2 rotation
curve as measured by \citet{Eilers2019}, together with a few other
measurements such as the total mass of the MW estimated by
\citet{Callingham2019} and the vertical force above the disc at the
Solar location given by \citep{Kuijken1991}. We found that a
contracted NFW DM halo model provides an excellent global fit to the
MW data (see Figure~\ref{fig:MW_fit_data}) and that it determines the
following properties for the MW components (see
Figure~\ref{fig:MW_fit_corner_plot}):
\begin{itemize} 
\item The Galactic DM halo has a mass of
  $M_{200}^{\rm DM}=0.97_{-0.19}^{+0.24}\times10^{12}\Msun$ and
  concentration before baryon contraction of $9.4_{-2.6}^{+1.9}$. The
  concentration value is identical to the median halo
  mass--concentration relation predicted by $\Lambda$CDM, suggesting 
  that the MW formed in a halo of average concentration. 
    \item The MW has a total mass of $M_{200}^{\rm
        total}=1.08_{-0.14}^{+0.20} \times 10^{12}\Msun$, in good
      agreement with many recent measurements based on the
      \textit{Gaia} DR2 data. 
    \item The MW stellar mass is $M_{\star\ \rm
        total}=5.04_{-0.52}^{+0.43}\times10^{10}\Msun$, of which
      roughly $60\%$ is found in the thin disc, and $20\%$ each in the
      thick disc and the bulge. This corresponds to a bulge-to-total ratio
      of 0.2. 
    \item The DM density at the Solar position is $\rho_{\odot}^{\rm
        DM} = 8.8_{-0.5}^{+0.5}\times10^{-3}\Msun\pc^{-3} \equiv 0.33_{-0.02}^{+0.02}~\rm{GeV}~\rm{cm}^{-3}$. 
\end{itemize}

While the contracted halo is the physically relevant model for the
Galactic mass distribution, we have also fitted an (uncontracted) pure
NFW halo model, mainly motivated by previous studies which have made
this assumption.  We have found that the same data are also well fit
by the pure NFW halo profile but with very different properties from
the contracted NFW halo model. In particular, the pure NFW halo model
has a $20\%$ lower DM mass, a higher halo concentration,
$c=13.3_{-2.7}^{+3.6}$, and a more compact and $20\%$ larger stellar
mass than the contracted halo model (see
Figure~\ref{fig:MW_fit_corner_plot} for a detailed comparison between
the two models).

The current rotation curve data used for the fit show a preference for
the contracted halo model, which has two times higher maximum
likelihood than the uncontracted halo. However, the difference is not
large enough to rule out the pure NFW halo model. Measurements of
other quantities such as the MW stellar mass, total mass, escape
velocity, as well as of the stellar-to-halo mass relation, all show
better matches to the contracted halo model (see discussion in
Section~\ref{subsec:discussion:which_is_better}). However, the
uncertainties in current measurements are large enough that we cannot
unequivocally establish if the NFW model is inconsistent with the
observational data. More accurate data, particularly \textit{Gaia} measurements of the stellar disc and \HI{} measurements of the gaseous disc, should resolve this ambiguity.

\section*{Acknowledgements}
We are grateful to Thomas Callingham, Ashley Kelly, Zhaozhou Li, Joop Schaye and Volker Springel for helpful discussions and suggestions. 
\changed{We also thank the anonymous referee for their constructive comments.}
MC acknowledges support by the EU Horizon 2020 research and innovation programme under a Marie Sk{\l}odowska-Curie grant agreement 794474 (DancingGalaxies).
MC and CSF were supported by the Science and Technology 
Facilities Council (STFC) [grant number ST/I00162X/1, ST/P000541/1]
and by the ERC Advanced Investigator grant, 
DMIDAS [GA 786910].  AD is supported by a Royal Society University
Research Fellowship. 
FAG acknowledges financial support from CONICYT through the project FONDECYT Regular Nr. 1181264, and funding from the Max Planck Society through a Partner Group grant.

This work used the DiRAC facility, hosted by Durham University, managed by the Institute for
Computational Cosmology on behalf of the STFC DiRAC HPC Facility
(www.dirac.ac.uk). The equipment was funded by BEIS capital funding 
via STFC capital grants ST/K00042X/1, ST/P002293/1, ST/R002371/1 and
ST/S002502/1, Durham University and STFC operations grant 
ST/R000832/1. DiRAC is part of the National e-Infrastructure.

\vspace{-.3cm}
\bibliographystyle{mnras}
\bibliography{MW_halo}

\appendix

\section{The inferred MW mass profile}
\label{eppendix:mass_profile_table}

\begin{table}
    \centering
    \caption{ \changed{The spherically averaged \textit{enclosed} mass profile of our galaxy inferred in this work. The table gives the maximum likelihood value and the 68 percentile confidence regions for the enclosed stellar, $M_{\star}^{\rm MW}$, baryonic, $M_{\rm bar}^{\rm MW}$, DM, $M_{\rm DM}^{\rm MW}$, and total, $M_{\rm total}^{\rm MW}$, mass as a function of radial distance from the Galactic Centre, $r$. The baryonic mass includes stars, \HI{}, molecular gas and CGM (see Section \ref{sec:MW_components} for details). The results are for our physically motivated contracted halo model. We only fit data in the range, $r\in[4,220]\kpc$, and the values outside this interval are model extrapolations.} }
   
    \renewcommand{\arraystretch}{1.15} 
    \begin{tabular}{ @{} l l l l l @{} } 
        \hline\hline
        $r$ & $M_{\star}^{\rm MW}(<r)$  & $M_{\rm bar}^{\rm MW}(<r)$ & $M_{\rm DM}^{\rm MW}(<r)$ & $M_{\rm total}^{\rm MW}(<r)$ \\
        $[\kpc]$ & $[10^{10}\Msun]$  & $[10^{10}\Msun]$ & $[10^{10}\Msun]$ & $[10^{10}\Msun]$ \\
        \hline  \\[-.3cm]
        
        1  &  $0.59_{ -0.04 }^{ +0.03 }$   &  $0.59_{ -0.04 }^{ +0.03 }$   &  $0.21_{ -0.02 }^{ +0.03 }$   &  $0.80_{ -0.04 }^{ +0.04 }$ \\
  2  &  $1.39_{ -0.09 }^{ +0.07 }$   &  $1.40_{ -0.09 }^{ +0.07 }$   &  $0.65_{ -0.07 }^{ +0.09 }$   &  $2.04_{ -0.07 }^{ +0.07 }$ \\
  3  &  $2.04_{ -0.13 }^{ +0.10 }$   &  $2.07_{ -0.13 }^{ +0.11 }$   &  $1.18_{ -0.12 }^{ +0.15 }$   &  $3.26_{ -0.07 }^{ +0.07 }$ \\
  5  &  $3.06_{ -0.22 }^{ +0.19 }$   &  $3.19_{ -0.22 }^{ +0.19 }$   &  $2.45_{ -0.22 }^{ +0.27 }$   &  $5.64_{ -0.05 }^{ +0.07 }$ \\
  8  &  $4.07_{ -0.34 }^{ +0.29 }$   &  $4.41_{ -0.34 }^{ +0.30 }$   &  $4.59_{ -0.36 }^{ +0.43 }$   &  $8.99_{ -0.07 }^{ +0.09 }$ \\
 10  &  $4.45_{ -0.40 }^{ +0.34 }$   &  $4.93_{ -0.40 }^{ +0.34 }$   &  $6.06_{ -0.43 }^{ +0.51 }$   &  $10.98_{ -0.10 }^{ +0.12 }$ \\
 15  &  $4.88_{ -0.48 }^{ +0.40 }$   &  $5.67_{ -0.49 }^{ +0.41 }$   &  $9.72_{ -0.56 }^{ +0.65 }$   &  $15.38_{ -0.16 }^{ +0.18 }$ \\
 20  &  $4.99_{ -0.51 }^{ +0.42 }$   &  $6.03_{ -0.52 }^{ +0.44 }$   &  $13.31_{ -0.63 }^{ +0.73 }$   &  $19.34_{ -0.25 }^{ +0.29 }$ \\
 25  &  $5.03_{ -0.52 }^{ +0.43 }$   &  $6.26_{ -0.53 }^{ +0.46 }$   &  $16.80_{ -0.67 }^{ +0.80 }$   &  $23.06_{ -0.41 }^{ +0.47 }$ \\
 30  &  $5.03_{ -0.52 }^{ +0.44 }$   &  $6.42_{ -0.53 }^{ +0.46 }$   &  $20.18_{ -0.74 }^{ +0.94 }$   &  $26.60_{ -0.63 }^{ +0.74 }$ \\
 40  &  $5.04_{ -0.52 }^{ +0.44 }$   &  $6.67_{ -0.55 }^{ +0.48 }$   &  $26.6_{ -1.1 }^{ +1.4 }$   &  $33.3_{ -1.2 }^{ +1.5 }$ \\
 50  &  $5.04_{ -0.52 }^{ +0.44 }$   &  $6.89_{ -0.56 }^{ +0.50 }$   &  $32.5_{ -1.7 }^{ +2.2 }$   &  $39.4_{ -2.0 }^{ +2.4 }$ \\
 60  &  $5.04_{ -0.52 }^{ +0.44 }$   &  $7.12_{ -0.57 }^{ +0.52 }$   &  $38.0_{ -2.5 }^{ +3.2 }$   &  $45.1_{ -2.8 }^{ +3.4 }$ \\
 70  &  $5.04_{ -0.52 }^{ +0.44 }$   &  $7.36_{ -0.58 }^{ +0.55 }$   &  $43.1_{ -3.3 }^{ +4.2 }$   &  $50.5_{ -3.6 }^{ +4.5 }$ \\
 80  &  $5.04_{ -0.52 }^{ +0.44 }$   &  $7.63_{ -0.60 }^{ +0.57 }$   &  $47.9_{ -4.0 }^{ +5.2 }$   &  $55.6_{ -4.4 }^{ +5.6 }$ \\
 90  &  $5.04_{ -0.52 }^{ +0.44 }$   &  $7.90_{ -0.62 }^{ +0.60 }$   &  $52.4_{ -4.8 }^{ +6.3 }$   &  $60.3_{ -5.2 }^{ +6.8 }$ \\
100  &  $5.04_{ -0.52 }^{ +0.44 }$   &  $8.20_{ -0.64 }^{ +0.63 }$   &  $56.7_{ -5.5 }^{ +7.4 }$   &  $64.9_{ -6.0 }^{ +7.9 }$ \\
125  &  $5.04_{ -0.52 }^{ +0.44 }$   &  $9.0_{ -0.7 }^{ +0.7 }$   &  $ 66_{ -7 }^{ +10 }$   &  $ 75_{ -8 }^{ +11 }$ \\
150  &  $5.04_{ -0.52 }^{ +0.44 }$   &  $9.9_{ -0.8 }^{ +0.8 }$   &  $ 75_{ -9 }^{ +13 }$   &  $ 85_{ -10 }^{ +13 }$ \\
175  &  $5.04_{ -0.52 }^{ +0.44 }$   &  $10.9_{ -0.8 }^{ +0.9 }$   &  $ 83_{ -11 }^{ +15 }$   &  $ 94_{ -11 }^{ +16 }$ \\
200  &  $5.04_{ -0.52 }^{ +0.44 }$   &  $12.0_{ -0.9 }^{ +1.1 }$   &  $ 90_{ -12 }^{ +17 }$   &  $102_{ -13 }^{ +18 }$ \\
225  &  $5.04_{ -0.52 }^{ +0.44 }$   &  $13.1_{ -1.0 }^{ +1.2 }$   &  $ 96_{ -14 }^{ +19 }$   &  $109_{ -14 }^{ +20 }$ \\
250  &  $5.04_{ -0.52 }^{ +0.44 }$   &  $14.3_{ -1.1 }^{ +1.3 }$   &  $102_{ -15 }^{ +21 }$   &  $117_{ -16 }^{ +22 }$ \\
275  &  $5.04_{ -0.52 }^{ +0.44 }$   &  $15.6_{ -1.2 }^{ +1.5 }$   &  $108_{ -16 }^{ +23 }$   &  $124_{ -17 }^{ +24 }$ \\
300  &  $5.04_{ -0.52 }^{ +0.44 }$   &  $17.0_{ -1.3 }^{ +1.7 }$   &  $114_{ -17 }^{ +25 }$   &  $131_{ -19 }^{ +26 }$ \\
        \hline
    \end{tabular}
    
    \renewcommand{\arraystretch}{1.0}
    \label{tab:appendix_mass_profile}
\end{table}

\changed{The best parameter values of our MW mass model and their implementation in the \texttt{galpy} code \citep{Bovy2015} are publicly available at \href{https://github.com/MariusCautun/Milky_Way_mass_profile}{https://github.com/MariusCautun/Milky\_Way\_mass\_profile}. For ease of use, we also present the inferred profiles in Table~\ref{tab:appendix_mass_profile}.
}

\section{Uncertainties associated with fitting a MW model}
\label{appendix:good_fit}

\begin{figure*}
        \centering
        \includegraphics[width=\linewidth,angle=0]{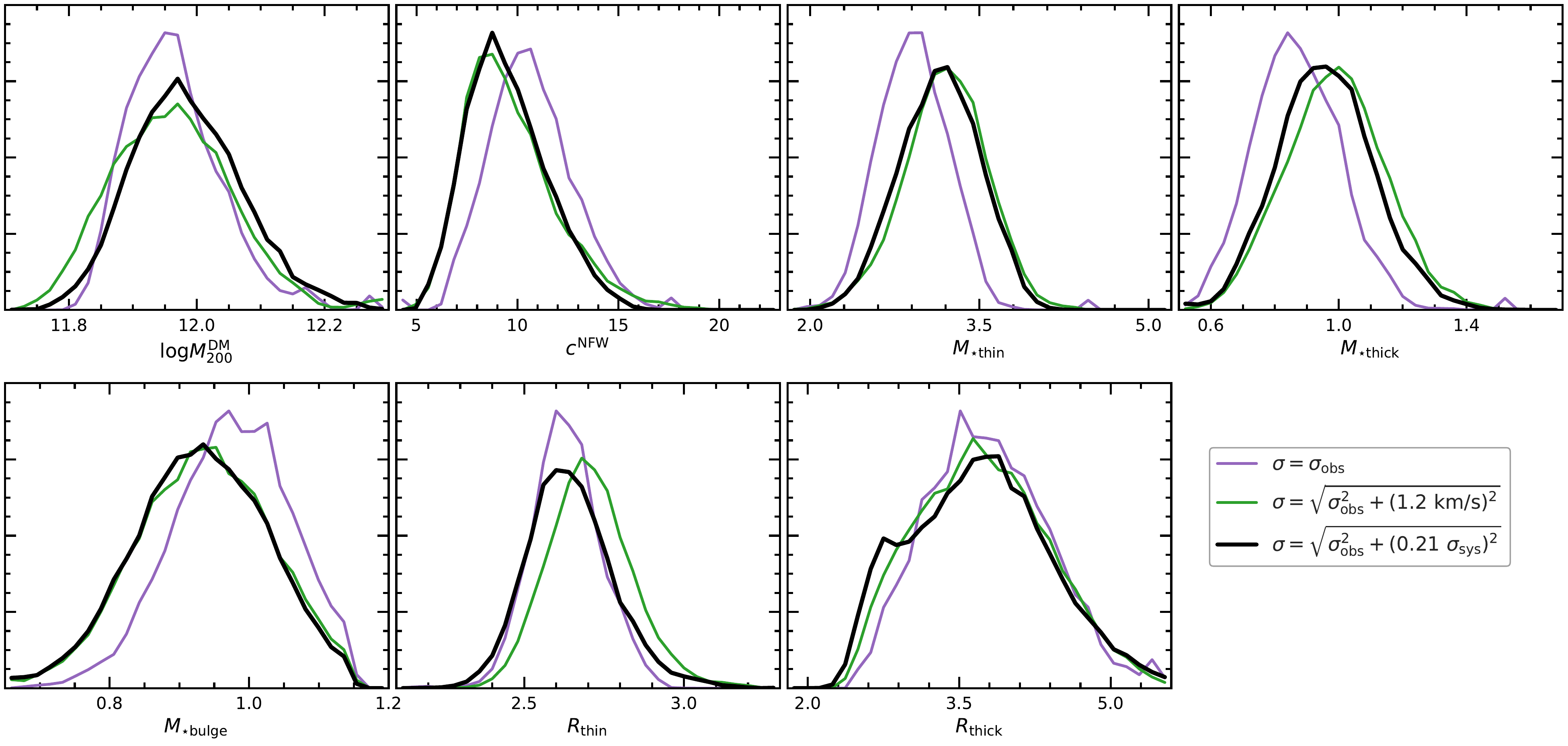}
        \vskip -.2cm
        \caption{ \changed{The marginalised PDF for the seven parameters used to describe the MW mass profile. Here we show the effect of adding model uncertainties to obtain a global fit  with a reduced $\chi^2$, close to unity. All approaches that include a model uncertainty (green and black curves) give roughly the same PDFs. The purple curve corresponds to finding the best fitting model in the absence of model uncertainties, i.e. using only the  \citet{Eilers2019} observational errors. In this case, the best fit model has a reduced $\chi^2\simeq2$. To obtain best fits with reduced $\chi^2\simeq1$, we add model uncertainties in two ways: i) by adding in quadrature a constant error of $1.2\kms$ to each data point (green curve), and ii) by adding in quadrature an uncertainty of $0.21\sigma_{\rm sys}$ (black curve; this is the default approach used in our analysis), where $\sigma_{\rm sys}$ is the systematic error associated with the  \citeauthor{Eilers2019} rotation curve. }
        }
        \label{fig:appendix_parameters_error_variation}
\end{figure*}

\changed{As we have argued in the main text, our goal is to find a global best fitting model of our galaxy. Models are often simplified versions of the complex processes affecting observations so there could be residual features in the data that manifest themselves as a reduced $\chi^2$ greater than 1. For example, structures in our galaxy, such as spiral arms, can affect the rotation curve measurements but they are not captured by the model studied here. As long as these fluctuations do no induce systematic changes in the rotation curve, we can account for them by adding an additional model uncertainty \citep[e.g. see][in the context of gravitational lensing]{Nightingale2018}. }

\changed{We use the data themselves to find the additional model uncertainty. To do so, we add a nuisance parameter to our model whose role is to quantify the size of the additional errors needed so that our best fit model returns a reduced $\chi^2$ of order unity. We have explored two different parametrisations for this nuisance parameter. First, we assume that this additional uncertainty is the same for all data points. Then, the error of each data point is the sum in quadrature of the \citet{Eilers2019} observational error, $\sigma_{\rm obs}$ and this constant value, $\sigma_0$, i.e. $\sigma=\sqrt{\sigma_{\rm obs}^2+\sigma_0^2}$. In a second approach, we assume that the additional uncertainty can vary between the data points. For example, this could be due to the various assumptions made when determining the rotation curve, with \citeauthor{Eilers2019} showing that these errors grow from $2$ percent for $R<15\kpc$ to ${\sim}10$ percent for $R>20\kpc$. Assuming that the \citeauthor{Eilers2019} systematic is the dominant source of deviation between our model and the data motives us to increase the rotation curve errors by adding a contribution proportional to the  \citeauthor{Eilers2019} systematic errors, $\sigma_{\rm sys}$. In this case, the total error for each data point is given by $\sigma=\sqrt{\sigma_{\rm obs}^2+(\mu\sigma_{\rm sys})^2}$, where $\mu$ denotes the nuisance parameter that we vary.}

\changed{We estimate the optimal value of the nuisance parameter by finding the value that maximises the likelihood. This works as follows. For a given model, the probability for an observation to have value $y$ is given by
\begin{equation}
    \frac{1}{\sqrt{2\pi}\sigma} \ \exp{ \frac{(y-y_{\rm model})^2}{2\sigma} } \;,
\end{equation}
where $y_{\rm model}$ denotes the model prediction and $\sigma$ the measurement plus the model uncertainty. Increasing the model uncertainty has two effects. A small value mostly affects the exponential term and can lead to a higher overall probability. However, too large a value for the model uncertainty hardly affects the exponential term, which will be close to unity, and will decrease the overall probability due to $\sigma$ appearing in the factor in front of the exponential. Thus, there is an optimal value of the model uncertainty that will maximise the probability.}

\changed{To find the optimal value of the model uncertainty, which is parametrised in terms of $\sigma_0$ or $\mu$, we run the MCMC algorithm using eight parameters: $\sigma_0$ (or $\mu$) plus the seven parameters described in Section \ref{sec:MW_model}. The likelihood is maximized for $\sigma_0=1.2\kms$ for the first parametrisation of the error and for $\mu=0.21$ for the second parametrisation (we actually obtain a range of values for $\sigma_0$ and $\mu$, but choose only the MLE). Then, we rerun the MCMC method but now keeping $\sigma_0$ (or $\mu$) fixed to their MLE values.} 

\changed{Figure \ref{fig:appendix_parameters_error_variation} shows the PDF for the seven MW parameters varied in our model. Each panel contains three curves corresponding to the case where we consider the original \citet{Eilers2019} errors and to the two cases where we add a model uncertainty to obtain a reduced $\chi^2$ close to unity. While we find some differences in the parameter distributions, these are rather small, especially for the two cases with model uncertainties. This indicates that our results are not sensitive to the approach employed to obtain a model that gives a good global fit to the data. In our analysis, we choose the second approach to increasing the error, i.e. $\sigma=\sqrt{\sigma_{\rm obs}^2+(\mu\sigma_{\rm sys})^2}$, since it predicts larger uncertainties at large distances, i.e. $R>20\kpc$. In that region, the \citet{Eilers2019} rotation curve shows a dip that is a few sigma away from the best fitting global model.}

\bsp	
\label{lastpage}

\end{document}